\documentclass[%
 aip,
 amsmath,amssymb,
 reprint,%
]{revtex4-1}
\usepackage{graphicx}
\usepackage{dcolumn}
\usepackage{bm}
\usepackage{color}
\usepackage[utf8]{inputenc}
\usepackage[T1]{fontenc}
\usepackage{mathptmx}
\usepackage{multirow}
\usepackage{float}
\usepackage{titlesec}
\titleclass{\subsubsubsection}{straight}[\subsection]
\newcounter{subsubsubsection}[subsubsection]
\renewcommand\thesubsubsubsection{\thesubsubsection.\alph{subsubsubsection}}
\titleformat{\subsubsubsection}
 {\normalfont\normalsize\bfseries}{\thesubsubsubsection}{1em}{}
\titlespacing*{\subsubsubsection}
{0pt}{3.25ex plus 1ex minus .2ex}{1.5ex plus .2ex}

\begin{document}

\preprint{AIP/123-QED}

\title{Morphological and non-equilibrium analysis of coupled Rayleigh-Taylor-Kelvin-Helmholtz instability}

\author{Feng Chen}
 \thanks{Corresponding author: chenfeng-hk@sdjtu.edu.cn, shanshiwycf@163.com}
 \affiliation{School of Aeronautics, Shan Dong Jiaotong University, Jinan 250357, China.}

\author{Aiguo Xu}
 \thanks{Corresponding author: Xu\_Aiguo@iapcm.ac.cn}
\affiliation{Laboratory of Computational Physics, Institute of Applied Physics and Computational Mathematics, P. O. Box 8009-26, Beijing 100088, P.R.China}
\affiliation{Center for Applied Physics and Technology, MOE Key Center for High Energy Density Physics Simulations, College of Engineering, Peking University, Beijing 100871, P.R.China}

\author{Yudong Zhang}
\affiliation{School of Mechanics and Safety Engineering, Zhengzhou University, Zhengzhou 450001, P.R.China
}

\author{Qingkai Zeng}
 \affiliation{School of Aeronautics, Shan Dong Jiaotong University, Jinan 250357, China.}

\date{\today}

\begin{abstract}
In this paper, the coupled Rayleigh-Taylor-Kelvin-Helmholtz instability(RTI, KHI and RTKHI, respectively) system is investigated using a multiple-relaxation-time discrete Boltzmann model. Both the morphological boundary length and thermodynamic non-equilibrium (TNE) strength are introduced to probe the complex configurations and kinetic processes. In the simulations, RTI always plays a major role in the later stage, while the main mechanism in the early stage depends on the comparison of buoyancy and shear strength. It is found that, both the total boundary length $L$ of the condensed temperature field and the mean heat flux strength $D_{3,1}$ can be used to measure the ratio of buoyancy to shear strength, and to quantitatively judge the main mechanism in the early stage of the RTKHI system. Specifically, when KHI (RTI) dominates, $L^{KHI} > L^{RTI}$ ($L^{KHI} < L^{RTI}$), $D_{3,1}^{KHI} > D_{3,1}^{RTI}$ ($D_{3,1}^{KHI} < D_{3,1}^{RTI}$); when KHI and RTI are balanced, $L^{KHI} = L^{RTI}$, $D_{3,1}^{KHI} = D_{3,1}^{RTI}$, where the superscript, ``KHI (RTI) " , indicates the type of hydrodynamic instability.
It is interesting to find that, (i) for the critical cases where KHI and RTI are balanced, both the critical shear velocity $u_C$ and Reynolds number $Re$ show a linear relationship with the gravity/accelaration $g$; (ii) the two quantities, $L$ and $D_{3,1}$, always show a high correlation, especially in the early stage where it is roughly $0.999$, which means that $L$ and $D_{3,1}$ follows approximately a linear relationship. The heat conduction has a significant influence on the linear relationship. A second sets of findings are as below: For the case where the KHI dominates at earlier time and the RTI dominates at later time, the evolution process can be roughly divided into two stages. Before the transition point of the two stages, $L^{RTKHI}$ initially increases exponentially, and then increases linearly. Hence, the ending point of linear increasing $L^{RTKHI}$ can work as a geometric criterion for discriminating the two stages. The TNE quantity, heat flux strength $D_{3,1}^{RTKHI}$, shows similar behavior. Therefore, the ending point of linear increasing $D_{3,1}^{RTKHI}$ can work as a physical criterion for discriminating the two stages.
\end{abstract}

\maketitle

\section{\label{sec:level1} Introduction}

Hydrodynamic instabilities are prevalent in various natural and technological environments. The Rayleigh-Taylor instability (RTI) occurs at a perturbed interface when a heavy fluid is accelerated or supported by a light one in a force field. The Kelvin-Helmholtz instability (KHI) occurs when there is a tangential velocity difference between two fluids separated by a perturbed interface. In the nonlinear evolution of RTI, the KHI will develop as a secondary instability at the high-density spike tips where the velocity shear is strong, and forms the resulted classical mushroom structures. Similarly, in the nonlinear evolution of KHI, the secondary RTI starts to develop along the vortex arms, due to the centrifugal acceleration of the rotating KHI vortex, when the density variation of the two fluids is large enough \cite{Faganello2008}. In addition, for a more general case, the coupled RTI and KHI are prevalent in various real systems, such as in the atmosphere and oceans, air/fuel mixing in combustion chambers, the outer region of supernovae, and the compression of the fuel capsule in inertial confinement fusion. Thus, it is a practical and inevitable problem to understand the behaviors of coupled Rayleigh-Taylor-Kelvin-Helmholtz instability (RTKHI).

Extensive efforts have been devoted to experimental and computational studies of the phenomena. Lawrence et al. \cite{Lawrence1991} performed a stability analysis on the coupled instability by solving the Taylor-Goldstein equation, compared with experiments performed in mixing layer channels, and discovered the transformation from shear instability dominated flow to RTI dominated flow. A linear analysis of the hybrid KHI and RTI in an electrostatic magnetosphere-ionosphere coupling system was carried out by Yamamoto \cite{Yamamoto2008}. The combined RTI and KHI of two superimposed magnetized fluids in the presence of suspended dust particles had been investigated by Prajapati et al\cite{Prajapati2009}. Guglielmi et al. discussed the coupled RTKHI at the magnetopause\cite{Guglielmi2010}, and the possible geophysical applications to the theory (e.g., penetration of the solar plasma into the magnetosphere, excitation of global Pc5 oscillations) were indicated. Ye et al. \cite{Wang2010,Ye2011} investigated analytically the competitions between RTI and KHI in two-dimensional incompressible fluids within a linear growth regime. It is found that the competition between the RTI and the KHI is dependent on the Froude number, the density ratio of the two fluids, and the thicknesses of the density transition layer and the velocity shear layer. Mandal et al. \cite{Mandal2011} investigated the nonlinear evolution of two fluid interfacial structures such as bubbles and spikes arising due to the combined action of RTI and KHI. Olson et al. \cite{Olson2011} studied the coupled RTKHI in the early nonlinear regime through Large eddy simulation. Numerical simulations showed a complex and non-monotonic behavior where small amounts of shear in fact decrease the growth rate, and the physical origins of this non-monotonic behavior were investigated. Dolai et al. \cite{Dolai2016} investigated the effect of different dust flow velocities and two dimensional magnetic fields on the combined KHI and RTI of two superimposed incompressible dusty fluids. Vadivukkarasan et al. \cite{Vadivukkarasan2016,Vadivukkarasan2017} described the three-dimensional destabilization characteristics of cylindrical and annular interfaces under the combined action of RTI and KHI mechanisms, by introducing dimensionless numbers such as Bond number, inner and outer Weber numbers and inner and outer density ratio. Sarychev et al. \cite{Sarychev2019} found that an undulating topography on the interface coating/base material resulted from a combination of Rayleigh-Taylor and Kelvin-Helmholtz instabilities.

When studying the coupled instability, the Richardson number $Ri$ is often used to quantify this transition from KHI-like to RTI-like behavior, and has been discussed by many authors \cite{SNIDER1996,Akula2013,Akula2016,Akula2017,Finn2014}. Akula et al. identified a similar transition from KHI-like to RTI-like instability growth used a gas tunnel facility, determined a transitional Richardson number of $-1.5$ to $-2.5$ in an early research \cite{Akula2013}, and $-0.17$ to $-0.56$ in later experiments ($At$ from $0.035$ to $0.159$) \cite{Akula2017}. Finn \cite{Finn2014} performed an experimental study of the combined RTI and KHI at three different Atwood numbers ($0.05$, $0.971$, $0.147$), and found that the transition occurs between the values of $-0.25$ and $-1.0$.

Besides the Hydrodynamic Non-Equilibrium (HNE) behaviors, the most relevant Thermodynamic Non-Equilibrium (TNE) behaviors in various complex flow systems, including systems with hydrodynamic instabilities, are attracting more attention with time \cite{xu2015aps,xu2016,Xu2018-Chapter2}. When the TNE is very weak, the loss of considering TNE does not make much difference. But when the TNE is strong, the situation will be greatly different. For example, the existence of TNE directly affects the density, temperature and pressure, as well as the magnitude and direction of flow velocity. Without considering TNE, the density, flow velocity, temperature and pressure given will have a significant deviation \cite{Lin-2019-CNF}. The existence of TNE is the underlying cause of heat flux and viscous stress. If insufficiently considered (considered only the linear response part), the amplitude of heat flow and stress obtained may be too large\cite{Gan2018}. When the strength of TNE beyond some threshold values, it may change the directions of heat flow and stress. If not sufficiently considered (considered only the linear response part), the resulting stress and heat flow, even in the wrong directions, may be obtained \cite{ZYD-2018}. In addition, in phase separation system, both the mean TNE strength \cite{gan2015} and the entropy production rate \cite{SoftMatter2019} increase with time in the spinodal decomposition stage and decrease with time in the domain growth stage, therefore, both the peak value points of the mean TNE strength and the entropy production rate can work as physical criteria to discriminate the two stages \cite{gan2015,xu2016,SoftMatter2019}. In system with combustion, the TNE behaviors help to better understand the physical structures of the von Neumann peak and various non-equilibrium detonation \cite{xu2015aps,yan2013,xu2015pre,Lin2016CNF,zhang2016CNF,Lin-2018-CAF,Lin-2017-SR,Xu-2018-FoP,Lin-2019-CNF}. For droplet collision, TNE behavior can be used to identify the different stages of the collision process and to distinguish different types of collision \cite{ZYD-FoP2020}. In system with RTI, the TNE behaviors around interfaces have been used to physically identify and distinguish various interfaces and design relevant interface-tracking schemes. All the TNE kinetic modes become stronger as the compressibility increases. Some TNE kinetic modes remain always in small amplitudes. With increasing the compressibility, more observable TNE kinetic modes appear for given observation precision\cite{Lai2016,Xu2018-Chapter2}. The correlation between the mean density nonuniformity and mean TNE strength is almost 1. The correlation between the mean temperature nonuniformity and mean Non-Organized Energy Flux (NOEF) is almost 1, and the correlation between the mean flow velocity nonuniformity and the mean Non-Organized Momentum Flux (NOMF) is also high, but generally less than 1 \cite{Chen2016,Xu2018-Chapter2}. The TNE effect helps to understand the effect of system dispersion (desctibed by $Kn$ number) on its kinetic behaviors\cite{Ye2020}. In system with pure KHI, via some defined TNE quantity, for example, the heat flux intensity, we can observe simultaneously the density interface and temperature interface so that we can investigate simultaneously the material mixing and energy mixing in the KHI evolution \cite{ZYD-2018,Gan2019}. The TNE behaviors were used to better understand the mixing entropy in multi-component flows\cite{Lin2017PRE}. In this work, we will show that, the endpoint of the NOEF, one of the various TNE effects, intensity linear growth stage can be used as a physical criterion from KHI-dominated to RTI-dominated in RTKHI coexistence systems.

Since the traditional hydrodynamic model is incapable of capturing the TNE behaviors, the above TNE combined by HNE studies resorted to a coarse-grained model derived from the Boltzmann equation, the discrete Boltzmann model (DBM) \cite{xu2015aps,xu2016,Xu2018-Chapter2}, developed from the well-known lattice Boltzmann method
\cite{ss2001,ss1992,xu2012,Yeoman1995PRL,Fang-Qian2004PRE,Qin2005PRE,Shan2016JFM,Shu2015JCP,Shu2007PRE,Zhong2012PRE,Zhong2015,Zhang-Qin2005JSP,xia2015,xia2010,SofoneaSpringer,Busuioc2019,Sofonea2020,ss2020,Qian2020,Shu2018,Zhong2020,shu2019,xuyu2020,Karlin2020,WangTan2019,Fei2019}. DBM is inspired by refining measurement step-by-step modeling scenario which is indicated by Chapman-Enskog (CE) multi-scale expansion. There are infinitely many specific ways to progressively refine the measurement, and CE describes one of them. In the case of the validity of CE theory, DBM construction can quickly confirm the necessary kinetic relations to be preserved by virtue of it, but, in principle, CE is only one of the candidate references. DBM, being different from the macroscopic fluid equations derived from CE in some aspects, shows physical advantages via the differences. The DBM has emerged as a feasible computational tool for describing the kinetic behaviors of complex systems. Besides recovering the macroscopic hydrodynamic equations in the continuum limit, DBM presents more kinetic information on the non-equilibrium effects which are generally related to some mesoscopic structures and/or kinetic modes \cite{xu2015aps,xu2016,Xu2018-Chapter2}. In 2012, Xu et al. \cite{xu2012} pointed out that, according to the non-equilibrium statistical physics, the nonconserved kinetic moments of $(f - f^{eq})$ can be applied to describe more specifically how the system deviates from its thermodynamic equilibrium state and to extract more specific information on the effects resulted from this deviation, where $f$ is the distribution function and $f^{eq}$ is the corresponding local equilibrium distribution function. $f$ and $f^{eq}$ share the same conserved kinetic moments, including the density, momentum and energy. Then, it was suggested to investigate the complex TNE behaviors in the phase space opened by the independent components of the nonconserved kinetic moments of $(f - f ^{eq})$ and its subspaces\cite{xu2015pre}. In the phase space opened by nonconserved kinetic moments and its subspaces, corresponding non-equilibrium strength was defined by means of the distance from the origin, and non-equilibrium state similarity and kinetic process similarity were defined by means of the reciprocal of the distance between two points \cite{xu2015pre,Xu2018-RGD31}. Via those concepts some previously unextractable information can be hierarchical, quantitative researched. These information work as a coarse-grained description of the TNE effects from various aspects \cite{xu2015aps,xu2016,Xu2018-Chapter2}. According to the extent of TNE that the model aims to describe, the DBM can be constructed in the levels of Navier-Stokes equations \cite{yan2013,lin2014,chen2014,gan2015,SoftMatter2019,xu2015pre,Lai2016,lai2018,Chen2016,Gan2019,
Lin2016CNF,zhang2016CNF,Lin2017PRE,Lin-2018-CAF,Lin-2017-SR,Xu-2018-FoP,Lin-2019-CNF,Ye2020}, Burnett equations\cite{Xu2018-Chapter2,ZYD-FoP2018,Gan2018,ZYD-2018},
etc. In terms of component number, besides the single fluid model, two or multiple-fluid DBM \cite{Lin2016CNF,Lin2017PRE,Lin-2017-SR,ZhangDJ2020-ESBGK-2F} for mixtures can be constructed according to the need. In terms of the collision model, besides the single-relaxation time model, multiple-relaxation time(MRT) DBM \cite{Chen2016} can be formulated. The DBM has brought significant new physical insights into various complex flows \cite{Xu2018-Chapter2}.

Besides by theory, results of DBM have been confirmed and supplemented by results of molecular dynamics \cite{kw2016,kw2017,kw2017b}, direct simulation Monte Carlo \cite{ZYD-2018,Meng2013JFM} and experiment \cite{Lin-2017-SR}. In recent years, DBM has been applied to investigate and has brought meaningful insights into various non-equilibrium behaviors in hydrodynamic instabilities. In 2018, we studied a coexisting system combined with RTI and Richtmyer-Meshkov instability (RMI) by using the DBM \cite{Chen2018}. It is found that, in both the pure RTI and pure RMI systems, the heat conduction plays a major role in influencing the correlation between the nonuniformity of hydrodynamic quantity (density, temperature or flow velocity) and non-equilibrium strength (mean TNE strength, mean NOEF or mean NOMF);  the correlation degree curves of the RTI system are relatively smooth, but in the RMI system there are many abrupt changes due to the existence and development of the shock wave. In the coexisting system combined with RTI, the parameter regions in which RMI and RTI dominate are given. The effects of gravity acceleration and Mach number on non-equilibrium were carefully studied.

In this paper, we investigate a coupled RTKHI system with the MRT DBM. Both the morphological boundary length and TNE strength are introduced to probe the complex configurations and kinetic processes. In order to conduct a systematic comparison, three cases are considered: (i) pure RTI, where relative velocity is set to zero; (ii) pure KHI, where acceleration is set to zero; and (iii) coupled RTKHI systems with various tangential velocities and accelerations.

The paper is organized as follows: Section II presents the MRT DBM with gravity and the morphological analysis technique. Systematic numerical simulations and analyses of pure RTI, pure KHI and coupled RTKHI systems are shown in Section III. A brief conclusion is given in Section IV.

\section{Brief review of Methodology}\label{Methodology}

According to the main strategy of the multiple-relaxation-time DBM scheme, the evolution of the discrete distribution function $f_{i}$ is given as
\begin{equation}
\frac{\partial f_{i}}{\partial t}+v_{i\alpha }\frac{\partial
f_{i}}{\partial
x_{\alpha }}=-\mathbf{M}_{il}^{-1}\hat{\mathbf{S}}_{lk}(\hat{f}_{k}-\hat{f}%
_{k}^{eq})-g_{\alpha }\frac{(v_{i\alpha }-u_{\alpha
})}{RT}f_{i}^{eq}\text{,}\label{dbm}
\end{equation}%
where the variable $t$ is the time, $x_{\alpha}$ is the spatial coordinate, $T$ is the temperature, $g_{\alpha}$ and $u_{\alpha}$ denote the macroscopic acceleration and velocity in the $x_{\alpha}$ direction, $v_{i\alpha}$ is the discrete particle velocity, $i=1$, $\ldots$, $N$, and the subscript $\alpha$ indicates the $x$, $y$, or $z$ component. $f_{i}$ and $\hat{f}_{i}$ ($f_{i}^{eq}$ and $\hat{f}_{i}^{eq}$) are the particle (equilibrium) distribution functions in velocity space and kinetic moment space (KMS), respectively; the mapping between moment space and velocity space is defined by the linear transformation $M_{ij}$, i.e., $\hat{f}_{i}=M_{ij}f_{j}$ and $f_{i}=M_{ij}^{-1}\hat{f}_{j}$. The matrix $\hat{\mathbf{S}}={\rm diag} (s_{1},s_{2},\cdots ,s_{N})$ is the diagonal relaxation matrix.

Here, the following two-dimensional discrete velocity model is used:
\begin{align}
\left(v_{ix,}v_{iy}\right) =\left\{
\begin{array}{cc}
\mathbf{cyc}\!:c\left( \pm 1,0\right) , & \text{for }1\leq i\leq 4, \\
c\left( \pm 1,\pm 1\right) , & \text{for }5\leq i\leq 8, \\
\mathbf{cyc}\!:2c\left( \pm 1,0\right) , & \text{for }9\leq i\leq 12, \\
2c\left( \pm 1,\pm 1\right) , & \text{for }13\leq i\leq 16,%
\end{array}%
\right.  \label{dvm3}
\end{align}%
and $\eta_{i}=\eta_{0}$ for $i=1$, \ldots, $4$, and $\eta_{i}=0$ for $i=5 $, \ldots , $16$, which is introduced to control the specific-heat-ratio $\gamma$. \textbf{cyc} indicates the cyclic permutation, and $c$ and $\eta_{0}$ are two free parameters, which are adjusted to optimize the properties of the model.

The transformation matrix $M_{ij}$ and the corresponding equilibrium distribution functions $\hat{f}_{i}^{eq}$ in KMS are constructed according to the seven moment relations $\hat{\mathbf{f}}^{eq}=\mathbf{M} \mathbf{f}^{eq}$. The details are as follows,
\begin{subequations}
\begin{equation}
\rho =\sum f_{i}^{eq},
\end{equation}%
\begin{equation}
\rho u_{\alpha }=\sum f_{i}^{eq}v_{i\alpha },
\end{equation}%
\begin{equation}
\rho \left( bRT+u_{\alpha }^{2}\right)/2 =\sum f_{i}^{eq}\left(
v_{i\alpha }^{2}+\eta _{i}^{2}\right)/2 ,
\end{equation}%
\begin{equation}
P\delta _{\alpha \beta }+\rho u_{\alpha }u_{\beta }=\sum
f_{i}^{eq}v_{i\alpha }v_{i\beta },
\end{equation}%
\begin{equation}
\rho \left[ \left( b+2\right) RT+u_{\beta }^{2}\right] u_{\alpha
}/2=\sum f_{i}^{eq}\left( v_{i\beta }^{2}+\eta _{i}^{2}\right)
v_{i\alpha }/2,
\end{equation}%
\begin{equation}
\rho \left[ RT\left( u_{\alpha }\delta _{\beta \chi }+u_{\beta
}\delta _{\alpha \chi }+u_{\chi }\delta _{\alpha \beta }\right)
+u_{\alpha }u_{\beta }u_{\chi }\right] =\sum f_{i}^{eq}v_{i\alpha
}v_{i\beta }v_{i\chi },
\end{equation}%
\begin{eqnarray}
&&\rho \left\{ \left( b+2\right) R^{2}T^{2}\delta _{\alpha \beta
}+\left[\left( b+4\right) u_{\alpha }u_{\beta }+u_{\chi }^{2}\delta _{\alpha \beta }%
\right] RT \right\}/2  \notag \\
&&+ \rho u_{\chi }^{2}u_{\alpha }u_{\beta }/2=\sum f_{i}^{eq}\left( v_{i\chi }^{2}+\eta _{i}^{2}\right) v_{i\alpha
}v_{i\beta }/2,
\end{eqnarray}%
\end{subequations}
where $\rho$, $T$, $P$\ are, respectively, the density, the temperature, and the pressure of gas. $\alpha, \beta, \chi=x, y$. By the Chapman-Enskog expansion on the two sides of the discrete Boltzmann equation, and modifying the collision operators of the moments related to energy flux, the Navier-Stokes equations with a gravity term for both compressible fluids and incompressible fluids can be obtained (See the previous work in Ref. \cite{Chen2016} for details).
\begin{subequations}
\begin{equation}
\frac{\partial \rho }{\partial t}+\frac{\partial (\rho u_{\alpha })}{%
\partial x_{\alpha }}=0\text{,}  \label{ns1}
\end{equation}%
\begin{eqnarray}
&&\frac{\partial (\rho u_{\alpha })}{\partial t}+\frac{\partial
\left( \rho
u_{\alpha }u_{\beta }\right) }{\partial x_{\beta }}+\frac{\partial P}{%
\partial x_{\alpha }}=-\rho g_{\alpha } \notag \\
&&+\frac{\partial }{\partial x_{\beta }}\left[\mu \left(\frac{%
\partial u_{\alpha }}{\partial x_{\beta }}+\frac{\partial u_{\beta }}{%
\partial x_{\alpha }}-\frac{2}{b}\frac{\partial u_{\chi }}{\partial x_{\chi }%
}\delta _{\alpha \beta }\right)\right]\text{,}
\label{ns2}
\end{eqnarray}%
\begin{eqnarray}
&&\frac{\partial e}{\partial t}+\frac{\partial }{\partial x_{\alpha}}\left[ (e+P)u_{\alpha }\right]=-\rho g_{\alpha }u_{\alpha} \notag \\
&&+\frac{\partial }{\partial x_{\beta }}\left[\lambda \frac{\partial T}{\partial x_{\beta}}+\mu \left(\frac{\partial u_{\alpha }}{\partial x_{\beta }}+\frac{\partial u_{\beta }}{\partial x_{\alpha }}-\frac{2%
}{b}\frac{\partial u_{\chi }}{\partial x_{\chi }}\delta _{\alpha
\beta }\right)u_{\alpha }\right]\text{,}  \label{ns3}
\end{eqnarray}%
\end{subequations}
where the coefficient of viscosity $\mu =\rho RT/s_{v}$ ($s_{v}=s_{5}=s_{6}=s_{7}$), and the heat conductivity $\lambda =(\frac{b}{2}+1)\rho R^{2}T/s_{T}$ ($s_{T}=s_{8}=s_{9}$).

Among these moment relations, only for the three, the definitions of density, momentum and energy, the equilibrium distribution function $f_{i}^{eq}$ can be replaced by the distribution function $f_{i}$. If we replace $f_{i}^{eq}$ by $f_{i}$ in right-hand-side (RHS) of any other required moment relations, the value of RHS will have a deviation from that of the left hand side. This deviation may work as a measure for the deviation of system from its equilibrium. This is the most basic idea for studying non-equilibrium \cite{xu2012}. In the MRT model, the deviation from equilibrium can be defined as $\Delta_{i}=\hat{f}_{i}-\hat{f}_{i}^{eq}={M}_{ij}(f_{j}-f_{j}^{eq})$. $\Delta_{i}$ contains the information of macroscopic flow velocity $u_{\alpha}$. On the basis, we replace $v_{i\alpha }$ by $v_{i\alpha}-u_{\alpha}$ in the transformation matrix $M_{ij}$, named $M_{ij}^{\ast }$. $\Delta _{i}^{\ast }=M_{ij}^{\ast}(f_{j}-f_{j}^{eq})$ is only the manifestation of molecular thermalmotion and does not contain the information of macroscopic flow. Corresponding to the simple definition of $\Delta _{i}^{\ast }$, we introduce some clear symbols as $\Delta_{2\alpha \beta }^{\ast}$, $\Delta_{(3,1)\alpha}^{\ast}$, $\Delta_{3\alpha \beta \gamma}^{\ast}$ and $\Delta_{(4,2)\alpha \beta }^{\ast}$, corresponding to $\Delta_{5,6,7}^{\ast}$, $\Delta_{8,9}^{\ast}$, $\Delta_{10,11,12,13}^{\ast}$ and $\Delta_{14,15,16}^{\ast}$, respectively. The subscript ``2" indicates the second-order tensor, and ``3,1" represents the first-order tensor contracted from a third-order tensor. Subscript ``xx" denotes the internal energy in the $x$ direction, ``xy" denotes the shear component, ``x" or ``xxx" denotes the flux in the $x$ direction. The TNE quantities are mostly around the interface where the gradients of macroscopic quantities are pronounced while approaching zeroes far away from the interface.

To provide a rough estimation of TNE, different TNE strength functions are defined\cite{Chen2016}, such as the globally averaged TNE strength $D_{TNE}$, Non-Organized Momentum Flux (NOMF) strength $D_{2}$ and Non-Organized Energy Flux (NOEF) strength $D_{(3,1)}$. The mathematic expressions are as below,
\begin{subequations}
\begin{equation}
D_{TNE}=\overline{d}=\overline{\sqrt{\Delta_{2\alpha \beta
}^{\ast 2}/T^{2}+\Delta_{(3,1)\alpha }^{\ast 2}/T^{3}+\Delta
_{3\alpha \beta \gamma }^{\ast 2}/T^{3}+\Delta_{(4,2)\alpha
}^{\ast 2}/T^{4}}} \text{,} \label{dtne}
\end{equation}%
\begin{equation}
D_{2}=\overline{d_{2}}=\overline{\sqrt{\Delta_{2\alpha
\beta}^{\ast 2}}} \text{,}\label{d2}
\end{equation}%
\begin{equation}
D_{(3,1)}=\overline{d_{3,1}}=\overline{\sqrt{\Delta_{(3,1)\alpha
}^{\ast 2}}} \text{.}\label{d31}
\end{equation}%
\end{subequations}

\begin{figure*}
\begin{center}
\includegraphics[bbllx=7pt,bblly=118pt,bburx=543pt,bbury=481pt,width=0.65\textwidth]{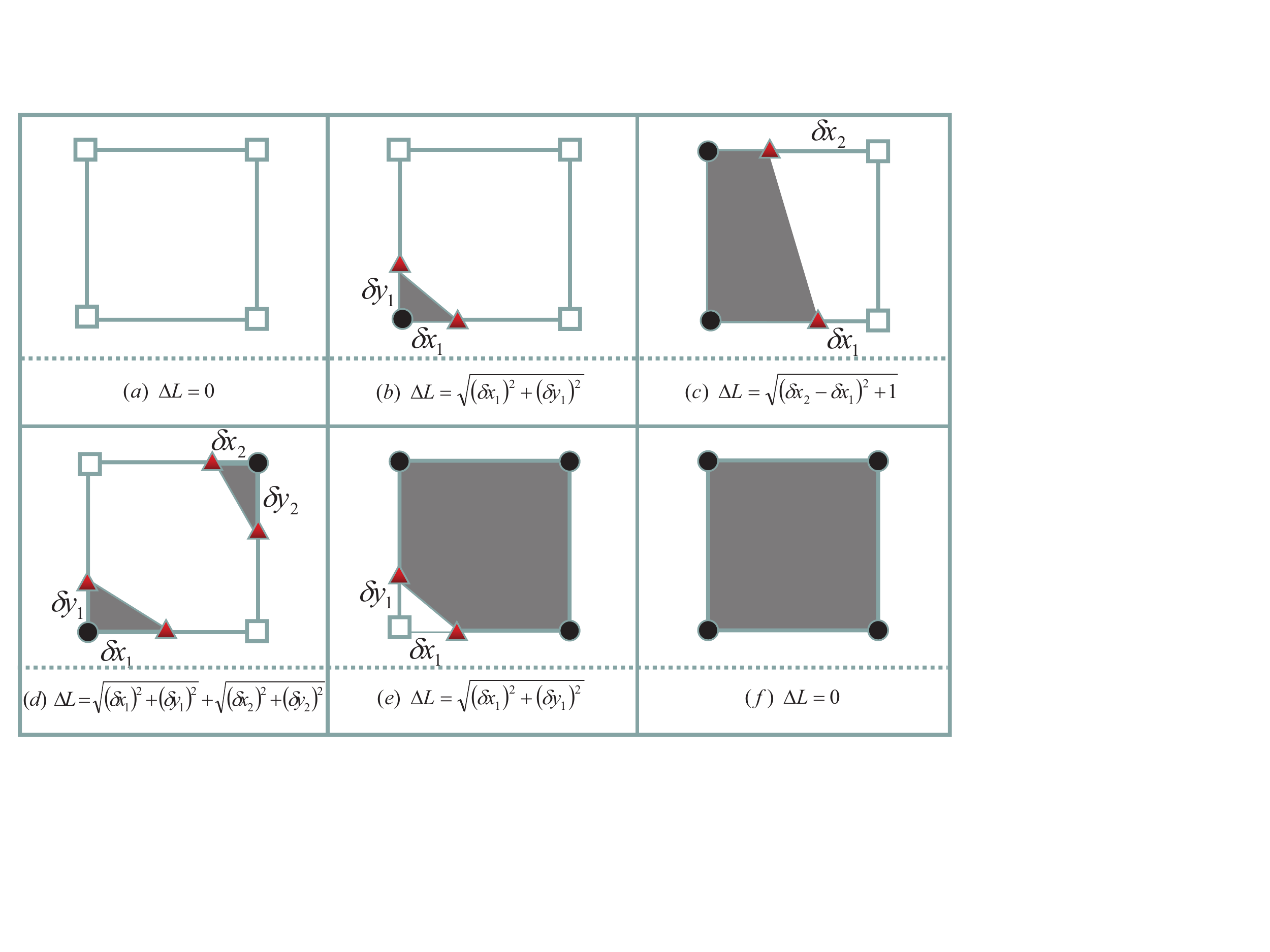}
\end{center}
\caption{The schematic diagram of morphological characterizations. Figures (a) to (f) present different pixel distributions, and the incremental calculations of boundary length in each case.}\label{fig1}
\end{figure*}

To quantitatively analyse the coupled RTKHI process, we resort to morphological analysis technique. Such a description has been well known in digital picture analysis, and successfully applied to characterize the density, temperature, pressure and particle velocity fields in shocked porous materials\cite{Xu2009}, to characterize the spinodal decomposition and domain growth processes in phase separation of multiphase flows\cite{Gan2011,Gan2012}, and to characterize the physical fields in evolution process of KHI system\cite{Gan2019}, etc. The basic idea of morphological analysis is as follows. A physical field $\theta(x,y)$ can be defined as two kinds of characteristic
regimes: When the physical quantity $\theta(x,y)$ is beyond the threshold value $\theta_{th}$, the grid node at position $(x,y)$ is regarded as a white (or hot) vertex, otherwise it is regarded as a black (or cold) one. A region with connected white or black nodes is defined as a white or black domain. In this way, the continuous image of physical field is converted into a Turing pattern, which is composed of only white and black pixels. For such a Turing pattern, a general theorem of integral geometry states that all the properties of a D-dimensional convex set satisfying motion invariance and additivity are contained in $D + 1$ Minkowski measures. To be specific, for a two dimensional field, the three Minkowski measures are the total fractional area $Area$ of the white regimes, the boundary length $L$ between the white and black regimes, and the Euler characteristic $\chi$ describing the connectivity of the domains. When the threshold $\theta_{th}$ is increased from the lowest to the highest values of $\theta(x,y)$ in the system, the white area $Area$ will decrease from 1 to 0; the boundary length $L$ first increases from 0, then arrives at a maximum value, and finally decreases to $0$ again. The Euler characteristic $\chi$ is defined as $\chi = N_W-N_B$, where $N_W$ ($N_B$) is the number of connected white (black) domains. The smaller the Euler characteristic $\chi$, the higher the connectivity of the structure with $\theta(x,y) \geq \theta_{th}$ or $\theta(x,y) < \theta_{th}$. A schematic diagram of morphological characterizations is shown in Fig. \ref{fig1}. Figures \ref{fig1}(a) to \ref{fig1}(f) represent different pixel distributions, and the incremental calculations of boundary length in each case are presented. The white squares (black circles) in the figure indicate that the values at these grid points are higher (lower) than the threshold value. The red triangles indicate that the values at the marked positions are equal to the threshold value, and the positions are determined by linear interpolation of the values on adjacent grid points.

In 2010 Xu, et al. \cite{Xu2010-SCPMA} suggested to use the phase space opened by the morphological quantities, $Area$, $L$ and $\chi$, to investigate the morphological property of two dimensional pattern. In such a morphological phase space, one point presents a complete description of the morphological characteristics of the pattern. The less the distance
\begin{equation}
Dist = \sqrt{(Area_2 - Area_1)^2 + (L_2 - L_1)^2 + (\chi_2 - \chi_1)^2}  \label{dist}
\end{equation}
between two points, 1 and 2, the higher the similarity of the two corresponding patterns in morphological characteristics. Therefore, the reciprocal of the distance $Dist$ between two points, $1/Dist$, can be defined as a pattern similarity $Sim$ in morphological characteristics, i.e.,
\begin{equation}
Sim = 1/Dist . \label{sim}
\end{equation}
When the distance $Dist = 0$, the similarity $Sim=\infty $, which means the morphological characteristics of the two patterns are the same. Go a further step, if the two patterns evolve with time from $t_1$ to $t_2$, then we can define a process similarity for the two pattern evolution processes during this period,\cite{Xu2010-SCPMA,Xu2011-CMA,Xu2016-SCPMA}
\begin{equation}
Sim_P = 1/Dist_P,\label{simp}
\end{equation}
where
\begin{equation}
Dist_P=\frac{1}{{\left( {{t_2} - {t_1}} \right)}}\int\limits_{{t_1}}^{{t_2}} {{Dist}(t){\rm{ }}dt} .\label{distp}
\end{equation}
The less the process similarity $Sim_P$, the larger the process difference $Dist_P$. These concepts and measuring methods have greatly promoted the development of physical cognition in related fields, such as phase separation, KH instability, shock wave kinetics of heterogeneous materials, etc.\cite{Xu2010-SCPMA,Xu2011-CMA,Xu2016-SCPMA}. In this work, we checked and found that, among the three morphological quantities, the boundary length $L$ perform the best in describing the evolution process of perturbed interface and the degree of material mixing. We will focus only on the behavior of $L$. Therefore, in this work,
\begin{equation}
Dist = \sqrt{(L_2 - L_1)^2}.  \label{dist-l}
\end{equation}

\section{Numerical Simulations}

The multiple-relaxation-time DBM model has been validated by some well-known benchmark tests, and satisfying agreements are obtained between the simulation results and analytical ones\cite{Chen2016, Chen2018}.
In this section, we investigate the coupled RTI and KHI system with this model; morphological and non-equilibrium analysis are also introduced to probe the complex process. We work in a frame where the gas constant $R=1$.

The initial macroscopic quantities in this study are given as follows:
\begin{eqnarray}
&T(y)=T_{u},\;\rho(y)=\rho _{u}\exp
(-g(y-y_{s})/T_{u}), \notag \\
&u_{x}(y)=u_{0}, \;u_{y}(y)=0, & \text{for } y\geq y_{s}, \notag
\\
&T(y)=T_{b},\;\rho(y)=\rho _{b}\exp (-g(y-y_{s})/T_{b}), \notag \\
&\;u_{x}(y)=-u_{0},\;u_{y}(y)=0, & \text{for } y < y_{s}, %
\label{rt}%
\end{eqnarray}%
where $y_{s}=40+2 \cos(0.1\pi x)$ denotes the interface with initial small perturbation. The computational domain is a two-dimensional box with height $H=80$ and width $W=20$. To be at equilibrium, the same pressure at the interface is required:
\begin{equation}
p_{0}=\rho_{u} T_{u}=\rho_{b} T_{b},  \label{rtp0}
\end{equation}
where $T_{u} < T_{b}$ and $\rho _{u} > \rho _{b}$. To have a finite width of the initial interface, all numerical experiments are performed by preparing the initial configuration plus a smooth interpolation between the two half-volumes. The initial temperature profile is therefore chosen to be
\begin{equation} \label{Eq:Ty}
T(y)=(T_{u}+T_{b})/2+(T_{u}-T_{b})/2 \times \tanh((y-y_{s})/w),
\end{equation}
where $w$ denotes the initial width of the interface. The initial density $\rho(y)$ is then fixed by the initial settings (Eqs. \eqref{rt}--\eqref{rtp0}) combined with the smoothed temperature profile. In the simulation, the bottom and the top boundaries are solid wall, the left and right boundaries are periodic boundary conditions. Time discretization is performed with a third-order Runge-Kutta scheme, and the fifth-order weighted essentially non-oscillatory (WENO) scheme is adopted for space discretization. The amplitude $A$ of pure RTI is often defined as half of the maximum distance between the the tips of bubble and spike. For the sake of contrast, the amplitudes of pure RTI, pure KHI and RTKHI systems are defined as half of the mixing width.

\subsection{Morphological analysis of pure RTI and pure KHI systems}

\begin{figure}[tbp]
\center\includegraphics*%
[bbllx=15pt,bblly=228pt,bburx=570pt,bbury=515pt,width=0.48\textwidth]{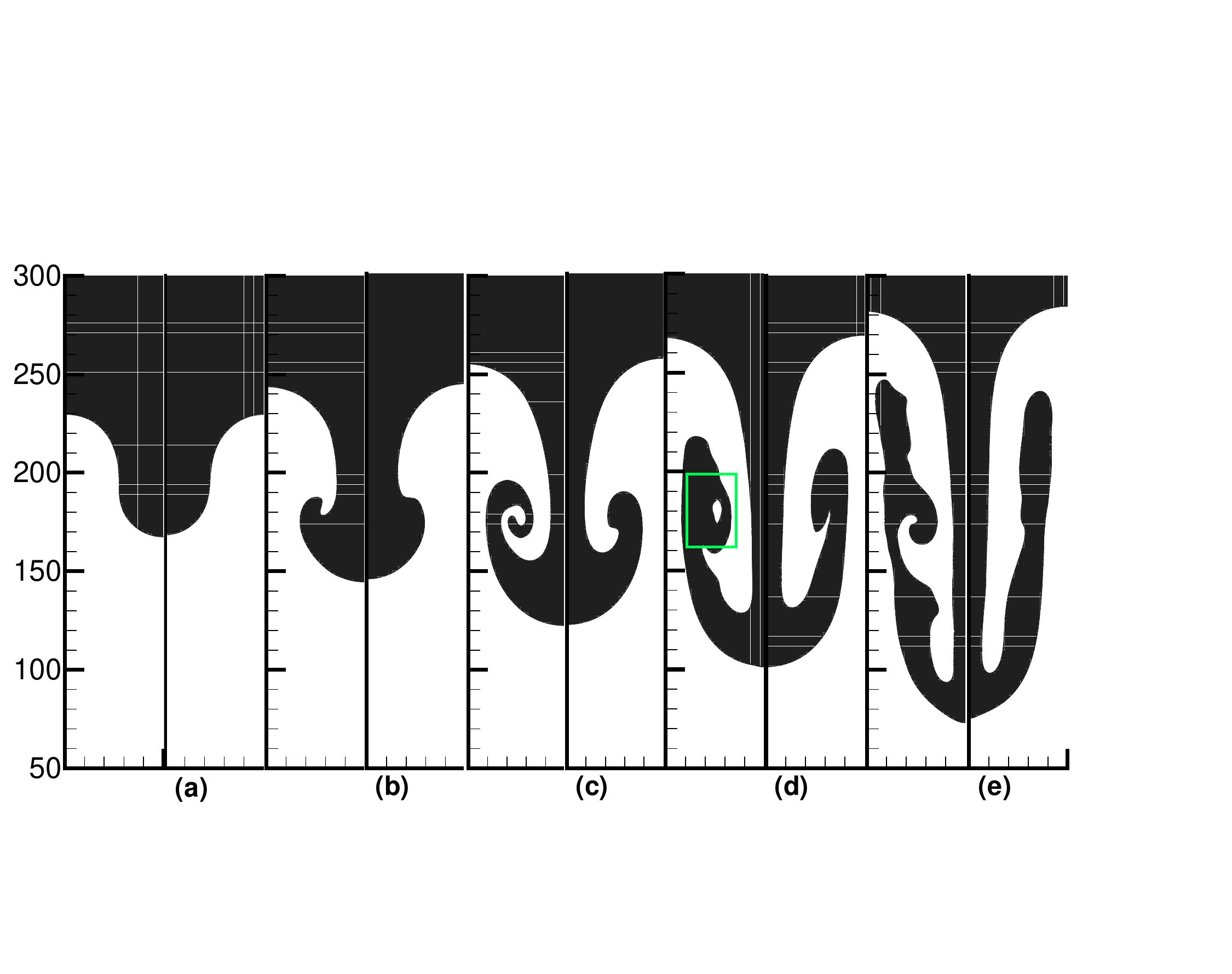}
\caption{Temperature Turing patterns of pure RTI ($g=0.005$, $u_{0}=0$, $T_{th}=1.0$). (a), (b), (c), (d) and (e) correspond to $t=100, 150, 200, 250, 300$. The left and right sides of each subgraph correspond to $s_{v}= 10^3$ and $s_{v}=10^2$, respectively.}\label{fig2}
\end{figure}

\begin{figure}[tbp]
\center\includegraphics*%
[bbllx=15pt,bblly=370pt,bburx=572pt,bbury=642pt,width=0.48\textwidth]{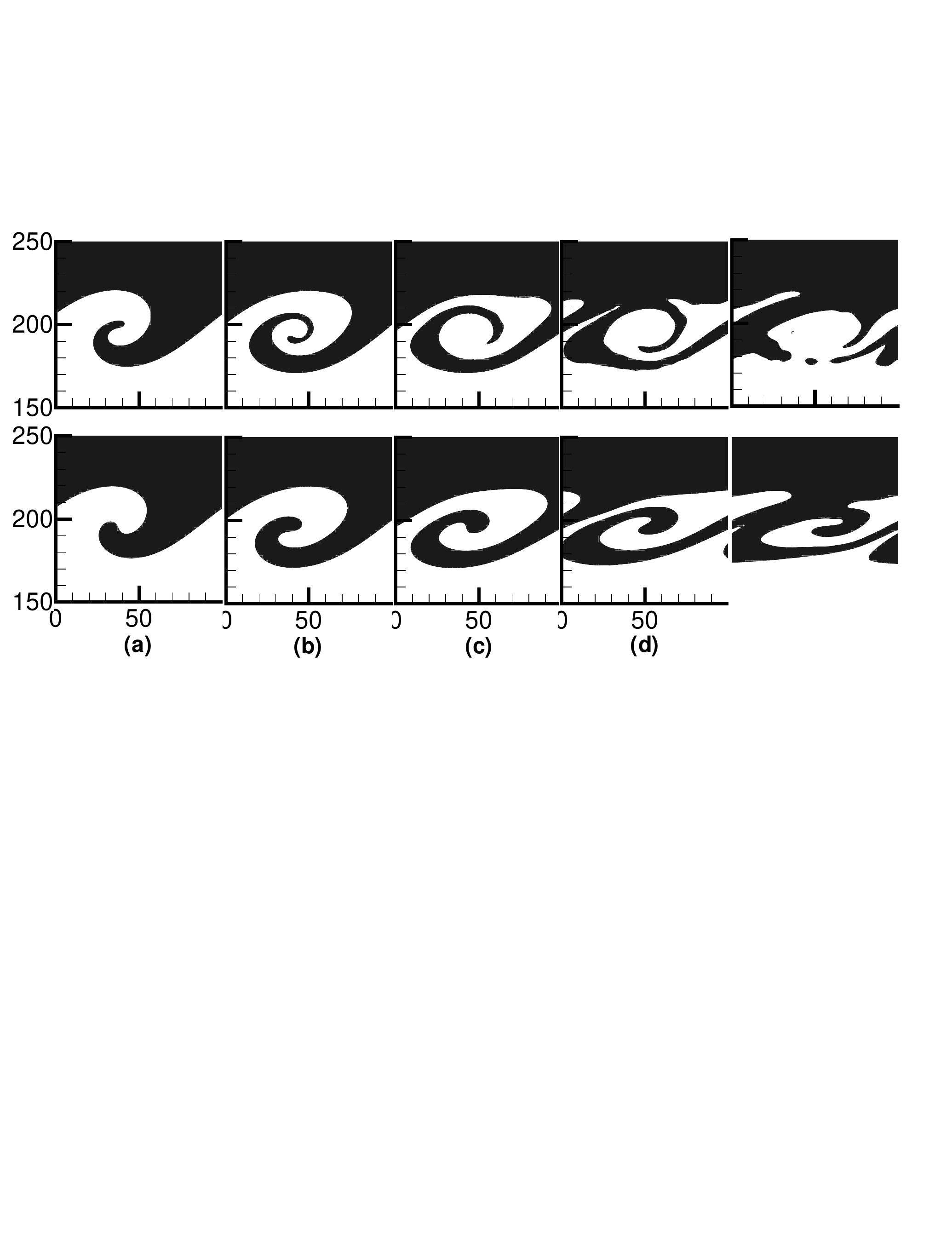}
\caption{Temperature Turing patterns of pure KHI ($g=0$, $u_{0}=0.125$, $T_{th}=1.0$). (a), (b), (c), (d) and (e) correspond to $t=100, 150, 200, 250, 300$. The upper and lower lines correspond to $s_{v}= 10^3$ and $s_{v}=10^2$, respectively.}\label{fig3}
\end{figure}

For comparison, we first simulate the pure RTI and the pure KHI systems. Figures \ref{fig2} and \ref{fig3} show the temperature Turing patterns of pure RTI ($g = 0.005$, $u_{0} = 0$) and pure KHI ($g = 0$, $u_{0} = 0.125$) at times $t = 100, 150, 200, 250, 300$, respectively. The threshold value is $T_{th}=1.0$. In Figure \ref{fig2}, the left and right sides of each subgraph correspond to $s_{v}= 10^3$ and $s_{v}=10^2$ (with other collision parameters being $10^{3}$), respectively. In Fig. \ref{fig3}, the upper and lower lines correspond to the simulation results of $s_{v}= 10^3$ and $s_{v}=10^2$ (with other collision parameters being $10^{3}$), respectively. The initial conditions and parameters are set as: $\rho_{b}=1 $, $T_{b}=1.4$, $\rho_{u}=2.33333$, $T_{u}=0.6$, $w=0.8$, $\gamma=1.4$, $c=1$, $\eta_{0}=3 $, $dx=dy=0.2$, $dt=10^{-3}$.

\begin{figure}[tbp]
\center\includegraphics*%
[scale=0.7,angle=0,bbllx=18pt,bblly=18pt,bburx=285pt,bbury=210pt\textwidth]{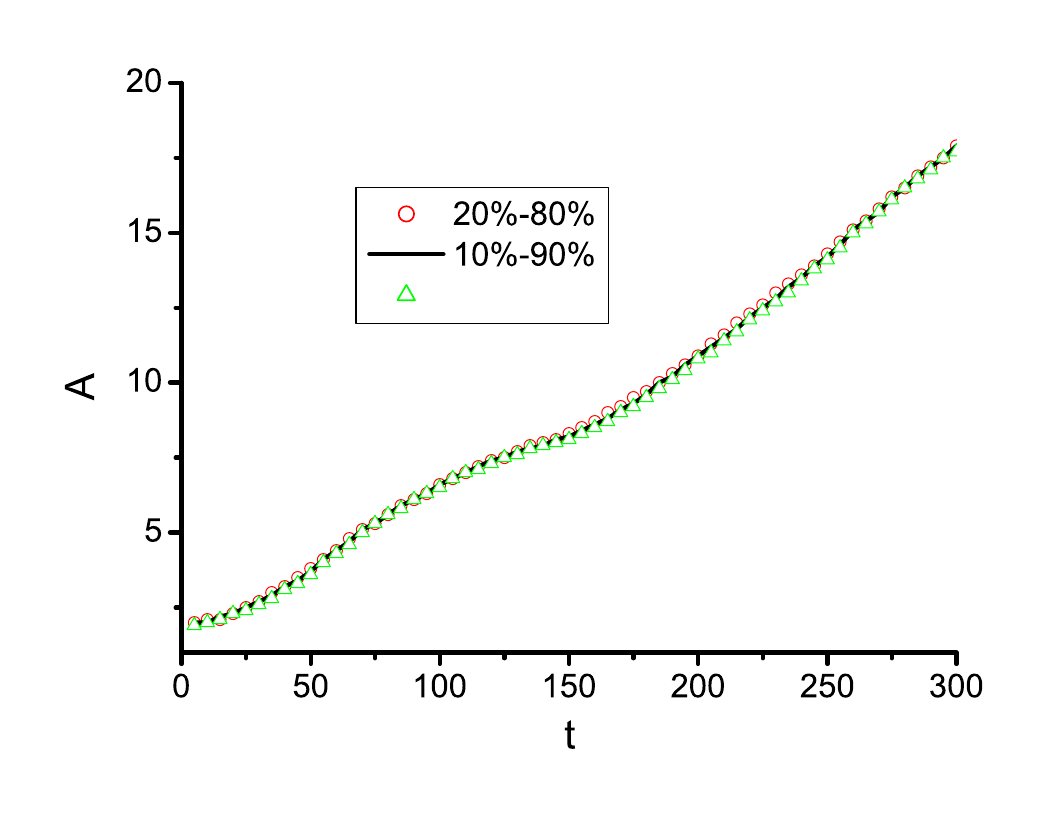}
\caption{Calculation of the amplitude $A$ for the case ($g=0.005$ and $u_{0}=0.1$).} \label{fig4}
\end{figure}

\begin{figure}[tbp]
\center\includegraphics*%
[bbllx=17pt,bblly=19pt,bburx=295pt,bbury=219pt,width=0.48\textwidth]{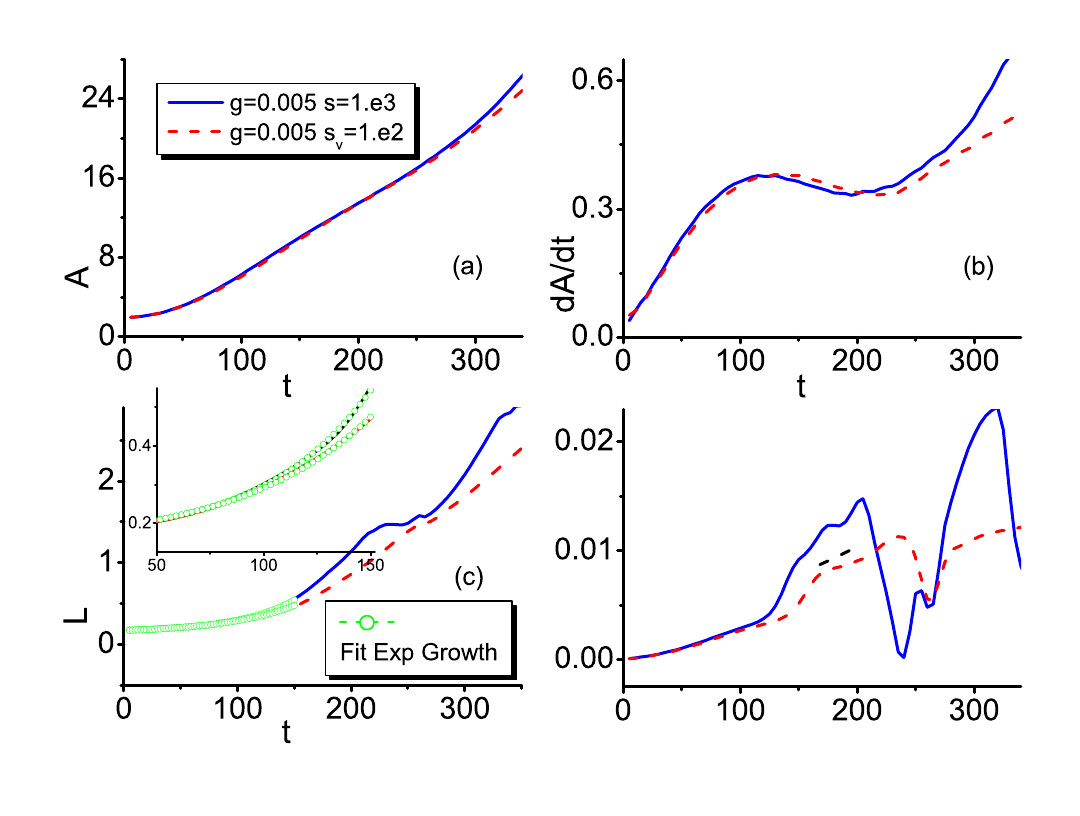}
\caption{The pure RTI, (a) amplitude $A$, (b) growth rate of amplitude $dA/dt$, (c) morphological boundary length $L$ of the temperature field, (d) growth rate of boundary length $dL/dt$.}\label{fig5}
\end{figure}

\begin{figure}[tbp]
\center\includegraphics*%
[bbllx=18pt,bblly=19pt,bburx=314pt,bbury=228pt,width=0.48\textwidth]{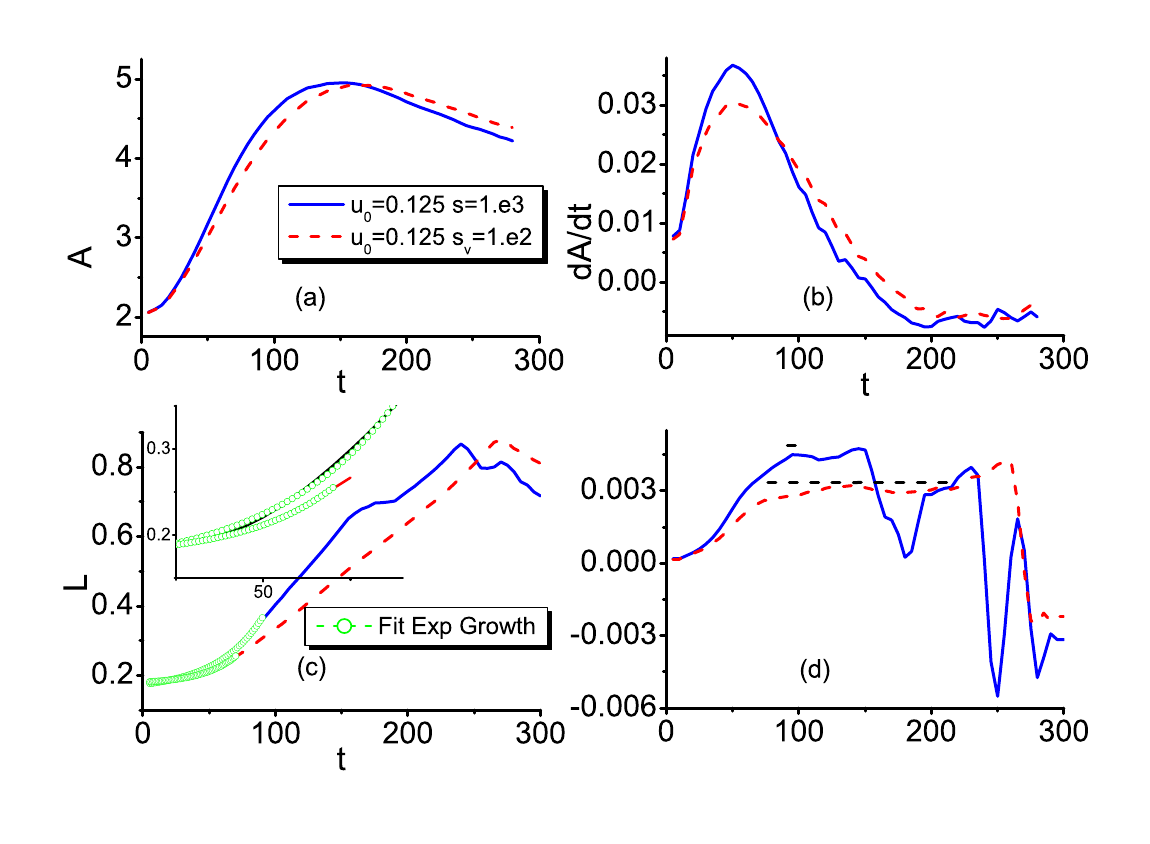}
\caption{The pure KHI, (a) amplitude (1/2 width of the mixing layer) $A$, (b) growth rate of amplitude $dA/dt$, (c) morphological boundary length $L$ of the temperature field, (d) growth rate of boundary length $dL/dt$.}\label{fig6}
\end{figure}

In many experimental studies of coupled RTKHI, $5\%$ and $95\%$ (or $10\%$ and $90\%$) volume fraction points (with respect to bottom stream fluid) are often chosen as the criterion to locate these upper and lower edges, respectively, in order to calculate the mixing width. This criterion has been justified by Snider and Andrews in 1994 and used in many work \cite{Akula2013,Akula2016,Akula2017}. In our numerical experiment, the temperature of the light fluid at the bottom is $1.4$, and the temperature of the heavy fluid at the top is $0.6$. The $\tanh$ function is used for smooth transition at the interface, as shown in Eq. \eqref{Eq:Ty}. So, in our morphological analysis, the range of temperature $T$ is $0.6 \le T \le 1.4$. Within this range, the $10\%$ and $90\%$ correspond to $T = 0.68$ and  $T = 1.32$, respectively, which are chosen as the criterion to locate these upper and lower edges to calculate the mixing width. In Fig. \ref{fig4}, the amplitude curves obtained by using different criterions for the case ($g=0.005$ and $u_{0}=0.1$) are shown. The black solid line corresponds to $10\%-90\%$ criterion, the red circles correspond to $20\%-80\%$ criterion, and the green triangles correspond to $5\%-95\%$ criterion. These results are in good agreement. Figures \ref{fig5} and \ref{fig6} show the time evolutions of amplitude $A$ and the morphological boundary length $L$ with threshold $T_{th}=1.0$ of the pure RTI and KHI, respectively.


In the evolution of pure RTI (as shown in Figs. \ref{fig2}, \ref{fig5}(a) and \ref{fig5}(b)), the perturbation amplitude is initially much smaller than the wave length, and the perturbation increases exponentially; as the light density (hot) and heavy density (cold) fluids gradually penetrate into each other, an obvious bubble-spike structure is gradually formed, the amplitude transitions from exponential growth to linear growth with time; subsequently, the top of the spike forms a mushroom-shaped structure due to the KHI, and the amplitude growth rate decreases slightly at this time; at a later time, extrusion from the two sides leads to the formation of secondary spikes, and the growth rate increases again (reacceleration stage). For fixed initial conditions and model parameters, high viscosity suppresses the development of RTI by inhibiting the development of KHI on the two sides of the spike.

In the evolution of pure KHI (as shown in Fig. \ref{fig3}), under the action of initial perturbation and tangential velocity, the perturbation gradually grows to a sinuous structure, and then a rolled-up vortex (Figs. \ref{fig3}(a)-\ref{fig3}(c)). In the final, the normal vortex structure collapses to nonregular structures and the system develops to the turbulent stage (Fig. \ref{fig3}(e)). Keeping other conditions and settings unchanged, the
larger the viscosity is, the weaker the KHI will be, and the later the vortex formulates and collapses.

Compared with the commonly used descriptions of amplitude and amplitude growth rate, the morphological boundary length $L$ can analyze the development of instability system from different angles.

i) We first compare the blue curves of amplitude and morphological boundary length of pure RTI in the case of $s_{v}=10^3$ in Fig. \ref{fig5}. The amplitude shows a smooth rising curve, and the amplitude growth rate increases
rapidly at $t = 250$, indicating that the system has entered the reacceleration stage. However, a small platform appears on the morphological boundary length $L$ at this time. The platform is a result of two factors: first, the continuous development of RTI increases $L$ (factor 1), and second, the destruction of the KHI vortex structure on both sides of the mushroom (marked by the green box in Fig. \ref{fig2}(d)) reduces $L$ (factor 2). The two factors
cancel each other. With the increasing viscosity, the boundary length $L$ decreases, and the degree of material mixing decreases. In addition, due to the suppression effect of viscosity, the effect of factor 2 just mentioned above is weakened, making the platform approximately disappear (red curve in Fig. \ref{fig5}(c)).

ii) In Fig. \ref{fig6}, before the fully developed turbulence stage, both the Mixing layer width and the boundary length $L$ of KHI increase at first, and then decreases, and the viscosity delays the appearance of the peak. When the Mixing layer width reaches the maximum, the vortex structure continues to develop, and the morphological boundary length $L$ continues to increase. The inflection points at which $L$ decreases (at times $t=240$ and
$270$) mean that the overall structure of the KHI vortex begins to break (as shown in Figs. \ref{fig3}(d) and \ref{fig3}(e)).

iii) Comparing the boundary length and its growth rate curves of RTI and KHI, it is found that, after the initial exponential growth stage, the boundary length of KHI has a constant velocity growth stage (marked with blue dotted lines), which is different from RTI, and the increase in viscosity helps to extend this process.

\subsection{Morphological analysis of Coupled RTKHI systems}

\begin{figure}[tbp]
\center\includegraphics*%
[bbllx=13pt,bblly=183pt,bburx=594pt,bbury=757pt,width=0.50\textwidth]{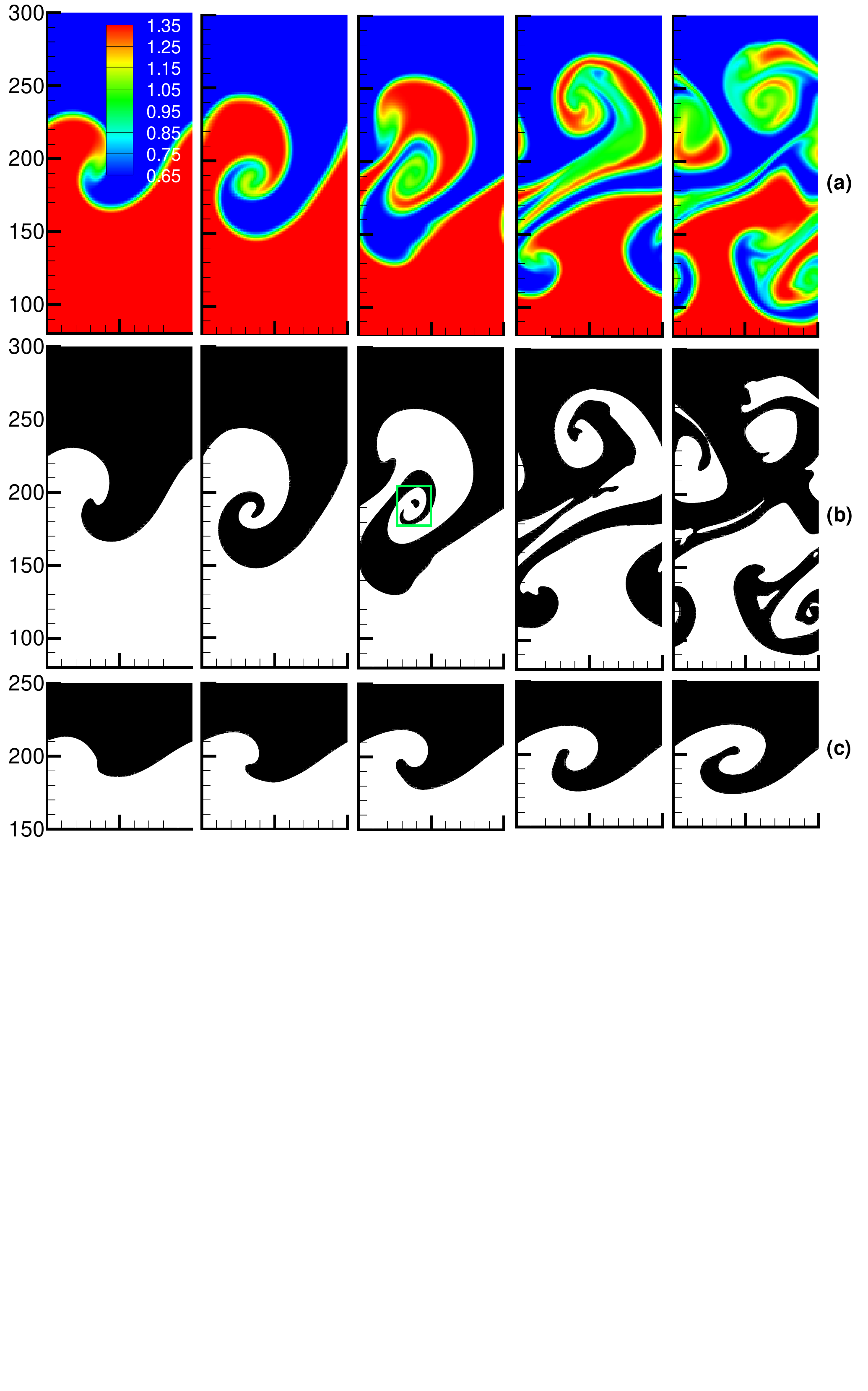}
\caption{A coupled RTKHI with $g=0.005$ and $u_{0}=0.05$. (a) and (b) are the temperature and the corresponding Turing pattern ($T_{th}=1.0$) of the RTKHI, respectively. (c) is temperature Turing pattern of pure KHI with $u_{0}=0.05$. The images from left to right correspond to $t=100,150,200,250,300$, respectively. All figures in the first line follow the same legend.}\label{fig7}
\end{figure}

\begin{figure}[tbp]
\center\includegraphics*%
[bbllx=13pt,bblly=209pt,bburx=594pt,bbury=757pt,width=0.50\textwidth]{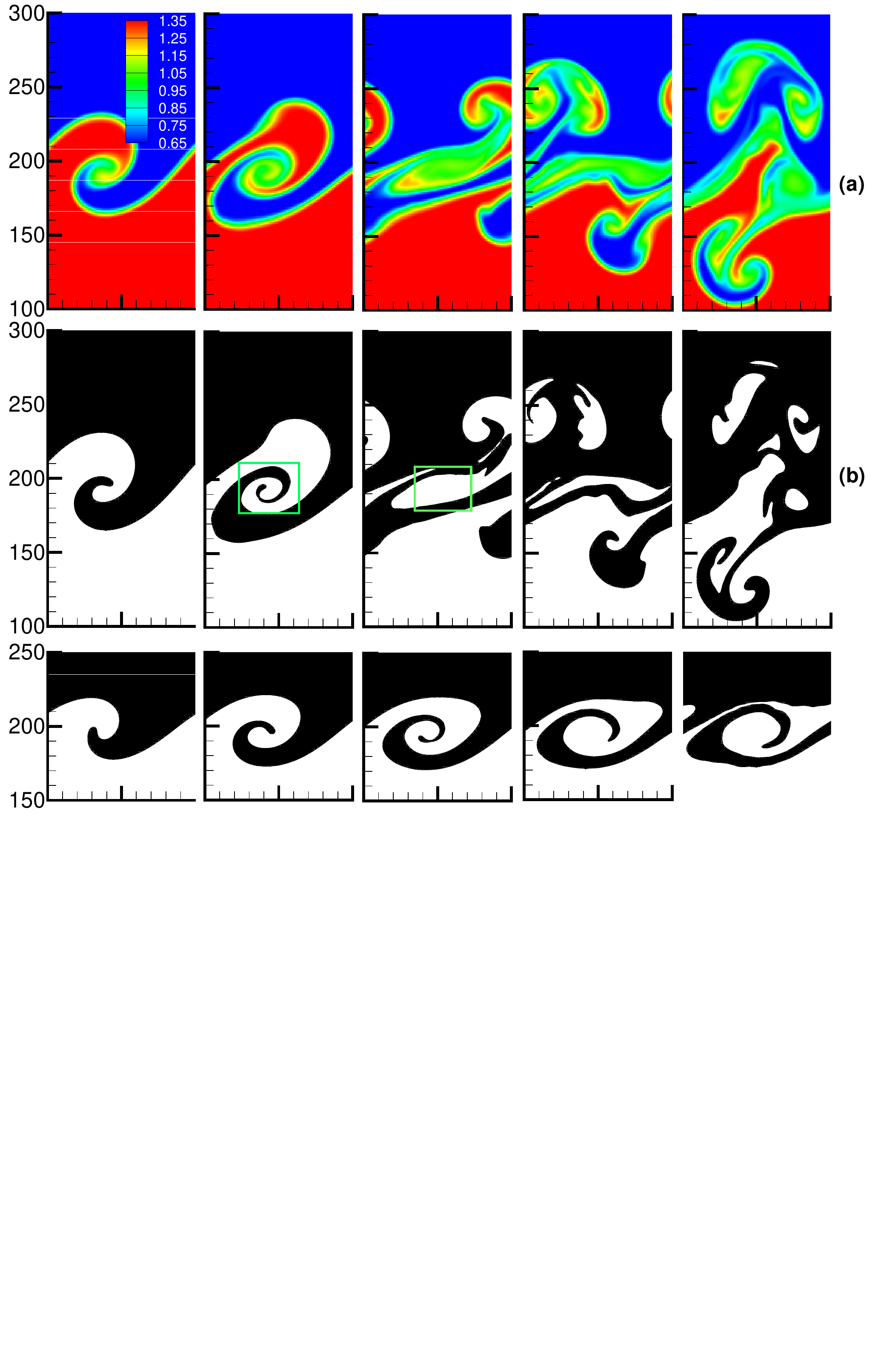}
\caption{A coupled RTKHI with $g=0.005$ and $u_{0}=0.1$. (a) and (b) are the temperature and the corresponding Turing pattern ($T_{th}=1.0$) of the RTKHI, respectively. (c) is temperature Turing pattern of pure KHI with $u_{0}=0.1$. The images from left to right correspond to $t=100,150,200,250,300$, respectively. All figures in the first line follow the same legend.}\label{fig8}
\end{figure}

\begin{figure}[tbp]
\center\includegraphics*%
[bbllx=13pt,bblly=209pt,bburx=594pt,bbury=757pt,width=0.50\textwidth]{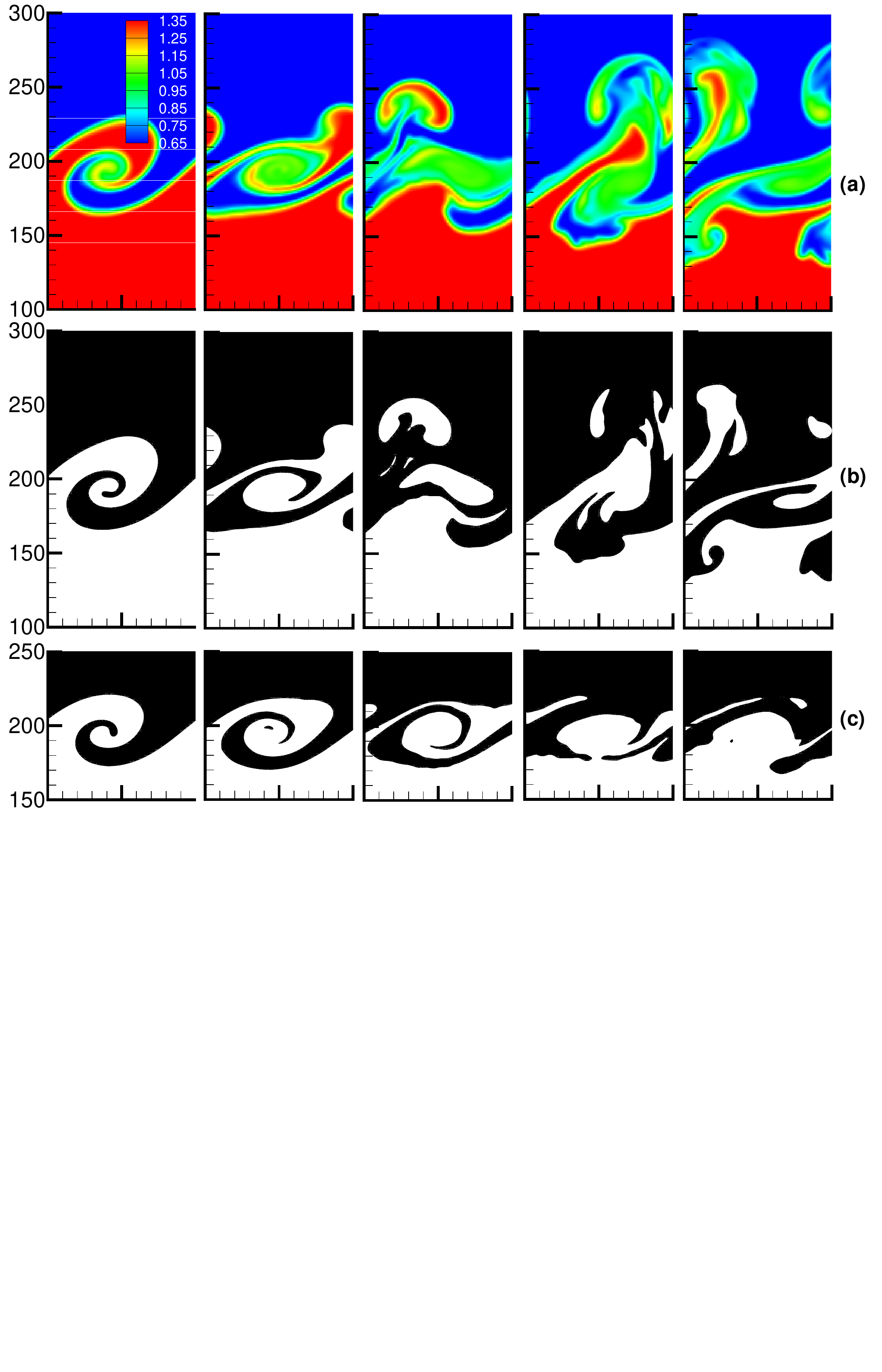}
\caption{A coupled RTKHI with $g=0.005$ and $u_{0}=0.15$. (a) and (b) are the temperature and the corresponding Turing pattern ($T_{th}=1.0$) of the RTKHI, respectively. (c) is temperature Turing pattern of pure KHI with $u_{0}=0.15$. The images from left to right correspond to $t=100,150,200,250,300$, respectively. All figures in the first line follow the same legend.}\label{fig9}
\end{figure}

In this section, we use morphological description to study the coupled RTKHI. Figures \ref{fig7}, \ref{fig8}, and \ref{fig9} show the temperature images and the corresponding Turing patterns of different RTKHI systems.


Figure \ref{fig7} corresponds to the case of $g=0.005$, $u_{0}=0.05$ (case 1). Compared with the pure RTI Turing patterns ($g = 0.005$, Fig.\ref{fig2}), and the pure KHI Turing patterns ($u_{0}=0.05$, Fig. \ref{fig7}(c)), it can be found that, the oblique and asymmetric bubbles, spikes and even mushroom-like structures are shown from beginning to end in Figs. \ref{fig7}(a) and \ref{fig7}(b), and the evolution of pure KHI in Fig. \ref{fig7}(c) lags far behind that in Fig. \ref{fig7}(b). Therefore, the RTI plays a major role, although KHI always exists in this system. The existence of shear velocity is mainly to destroy the symmetry of the RTI structure.

Figure \ref{fig8} corresponds to the case of $g=0.005$, $u_{0}=0.1$ (case 2). It can be found that, the interface and vortex growth of the RTKHI system, shown in Figs. \ref{fig8}(a) and \ref{fig8}(b), show similar magnitudes to those of the pure RTI, shown in Fig. \ref{fig2} and the pure KHI with $u_{0}=0.1$, shown in Fig. \ref{fig8}(c), before the time $t=150$. In this process, neither RTI nor KHI can be ignored. After the time $t=150$, the vortex structure of the system is destroyed (marked by the green boxes in Fig. \ref{fig8}(b)), and an asymmetric mushroom-like structure (representative structure of RTI) is also generated. At this time, RTI plays a major role.

Figure \ref{fig9} corresponds to the case of $g=0.005$, $u_{0}=0.15$ (case 3). The spanwise vortical structures which are characteristic of free shear flows are clearly seen at earlier time ($t <= 100$), so KHI plays a major role at the initial stage. As time progresses, more fluid is entrained in the vortical structure, and the secondary RTI develops rapidly along the vortex arms ($t=150$). Subsequently, the vortex structure of KHI is quickly stretched and diffused, and RTI drives the mixing layer growth ($t=200$). In the later turbulence stage, the combined effect of gravity acceleration and shear makes the interface more irregular ($t=250,300$).

\begin{figure*}
\begin{center}
\includegraphics[bbllx=15pt,bblly=15pt,bburx=305pt,bbury=230pt,width=0.8\textwidth]{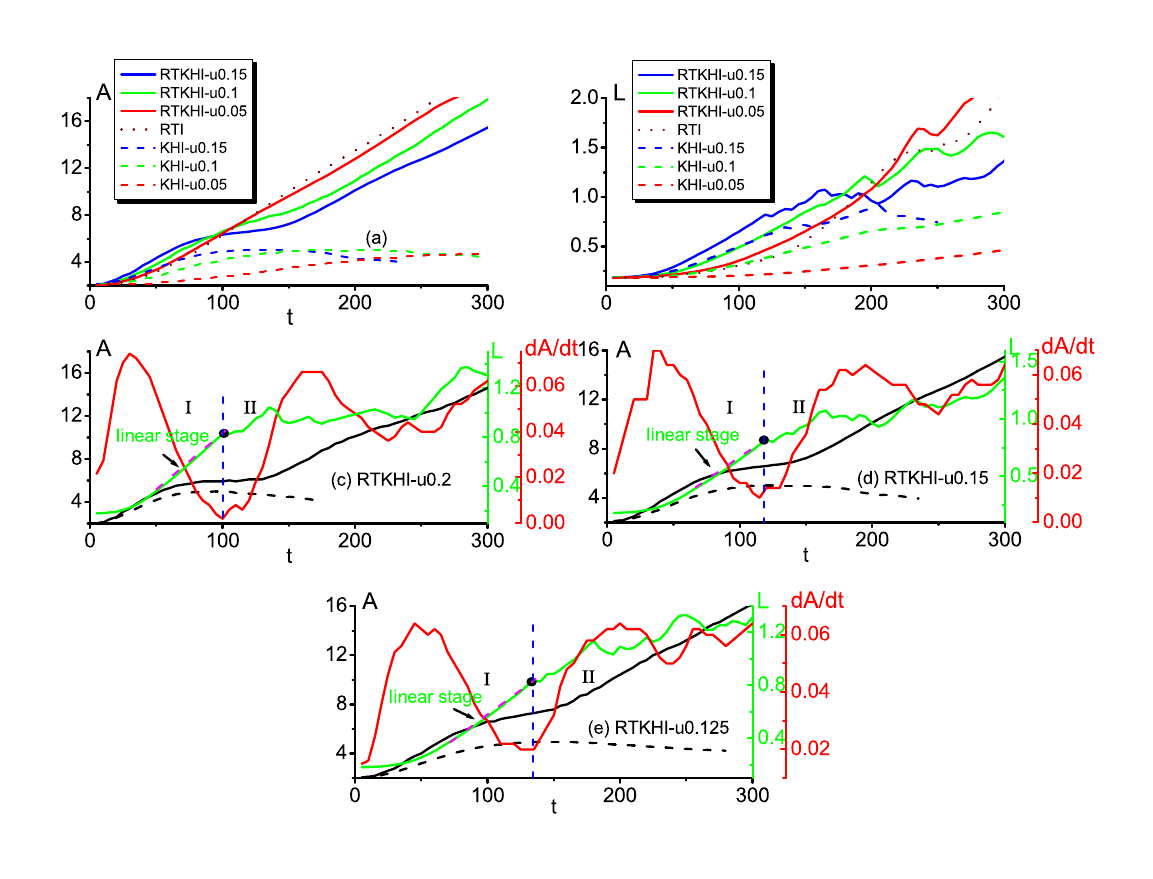}
\end{center}
\caption{Perturbation amplitude $A$, growth rate $d A/d t$, and morphological boundary length $L$ versus time $t$. For convenience of comparison, the results for the coupled RTKHI, pure RTI and pure KHI systems are presented in different form of graph. Figures (a) and (b) show the perturbation $A$ and morphological boundary length $L$ for various cases, including the coupled RTKHI, the pure RTI and pure KHI. In Figs. (c), (d) and (e),  the evolution of three quantities, $A$, $d A/d t$, and $L$, are shown in the same graph for one case of RTKHI system. The shear rates in (c), (d) and (e) are $u_{0}=0.2$, $u_{0}=0.15$, $u_{0}=0.125$, respectively. The black, red and green solid lines represent the amplitude, amplitude growth rate and boundary length of the RTKHI system, respectively. For comparison, the amplitude $A$ of the corresponding pure KHI is also plotted by black dashed line in Figs. (c), (d) and (e).}\label{fig10}
\end{figure*}
\begin{figure*}
\begin{center}
\includegraphics[bbllx=16pt,bblly=68pt,bburx=340pt,bbury=286pt,width=0.7\textwidth]{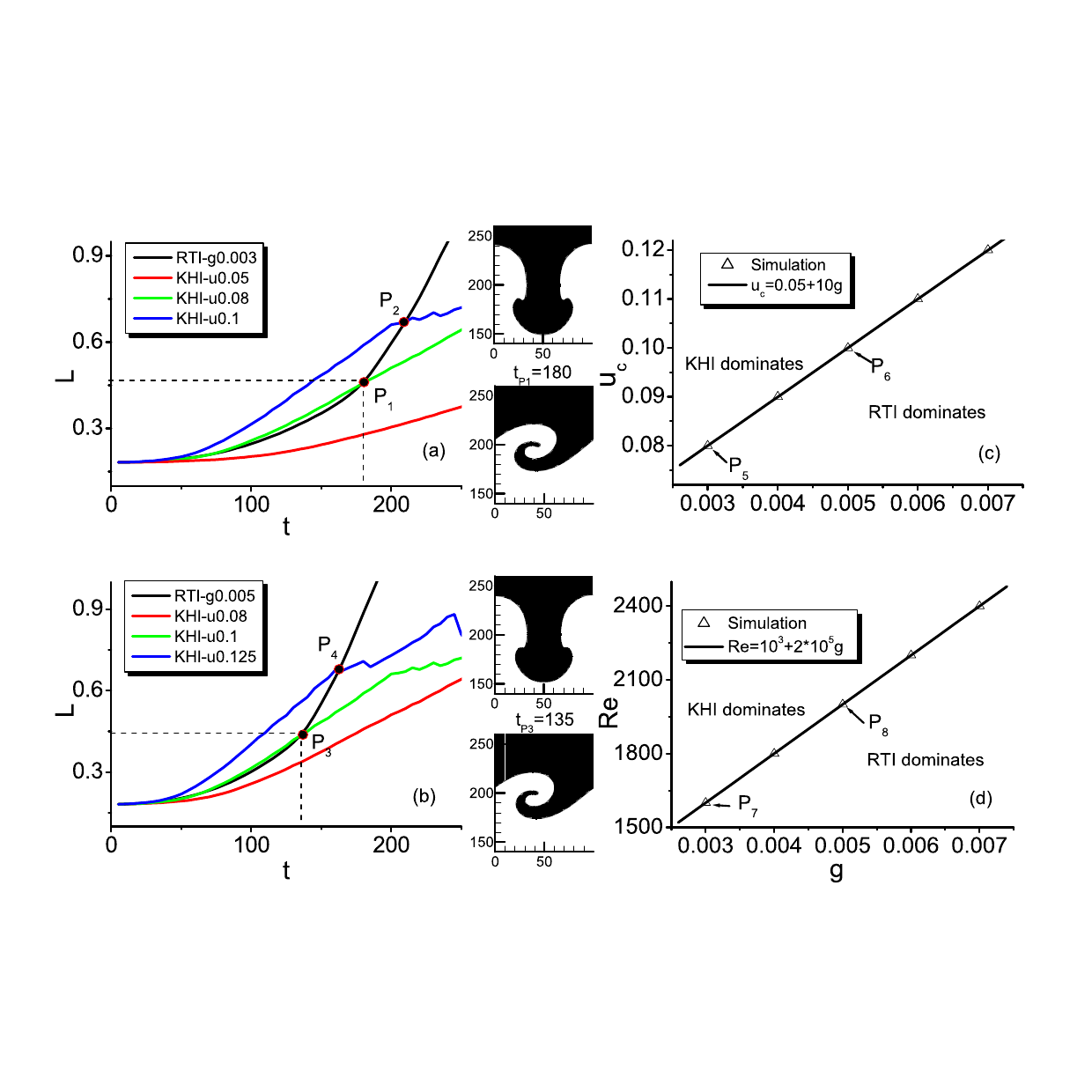}
\end{center}
\caption{Morphological analysis of the main mechanism in the early stage. Figures (a) and (b) show the comparison of boundary length $L$ between pure RTI and different pure KHI. The small insets in the middle column are the temperature Turing patterns of pure RTI (corresponding to black curves in (a) and (b)) and the corresponding pure KHI (corresponding to green curves in (a) and (b)) at the times marked by the intersection points, P$_1$ and P$_3$ in (a) and (b). (c) and (d) shows the linear relationships between gravity acceleration $g$ and critical shear velocity $u_C$, gravity acceleration $g$ and Reynolds number $Re$. The regions for KHI or RTI dominates are obtained.}\label{fig11}
\end{figure*}

Next, the focus of our work shifts to two judgments: i) Qualitatively speaking, for the early stage of the coupled RTKHI system, when the shear effect is relatively small, RTI dominates; when the shear is relatively large, KHI will play a major role. How to judge quantitatively ? ii) For the case where the KHI dominates at earlier time and the RTI dominates at later time, how to judge this transition point from KHI-like to RTI-like. It is important to quantify these transition points for different cases and determine any criterion for transition.

In Fig. \ref{fig10}, perturbation amplitude $A$, growth rate $d A/d t$, and morphological boundary length $L$ versus time $t$ are shown. For convenience of comparison, the results for the coupled RTKHI, pure RTI and pure KHI systems are presented in different form of graph. Figures \ref{fig10}(a) and \ref{fig10}(b) show the perturbation $A$ and morphological boundary length $L$ for various cases, including the coupled RTKHI, the pure RTI and pure KHI. In Figs. \ref{fig10} (c), (d) and (e), the evolution of three quantities, $A$, $d A/d t$, and $L$, are shown in the same graph for one case of RTKHI system. The shear rates in \ref{fig10}(c), (d) and (e) are $u_{0}=0.2$, $u_{0}=0.15$, $u_{0}=0.125$, respectively. The black, red and green solid lines represent the amplitude, amplitude growth rate and boundary length of the RTKHI system, respectively. For comparison, the amplitude $A$ of the corresponding pure KHI is also plotted by black dashed line in each of Figs. \ref{fig10}(c), (d) and (e). The superposition of shear on RTI increases the mixing width and boundary length at early time, while this effect is not observed at later time. The morphological boundary length curves are complicated and variable in the later time, and these complicated changes are due to the development and deformation of the vortex, and the coalescence of the hot and cold domains. The amplitude curves show that the system has a growth trend similar to pure RTI in the later stage, and verify the qualitative description mentioned above again. For the case where the KHI dominates at earlier time and the RTI dominates at later time, the evolution process can be roughly divided into two stages, that is, shear dominance stage and buoyancy dominance stage. The state of transition from shear dominance to buoyancy dominance is called the transition point. As can be seen from the figure, in the initial stage, $L^{RTKHI}$ initially increases exponentially, and then increases linearly (purple dashed lines). The ending point of linear increasing $L^{RTKHI}$ is approximately equal to the minimum point of amplitude growth rate $d A/d t$ of the RTKHI system, and the maximum point of amplitude $A$ of the corresponding pure KHI system. From this moment, the RTI begins to play a major role. This moment is referred to as the transition point, which is marked by a dashed vertical line and circle in each figure. Hence, the ending point of linear increasing $L^{RTKHI}$ can work as a geometric criterion for discriminating the two stages. The higher the shear rate is, the earlier the transition appears.

We also calculate the Richardson number $R_{i}$ corresponding to the transition point, based on the following equation\cite{Akula2013, Akula2017,Finn2014},
\begin{equation}
R_{i}=\frac{-g(\partial \rho /\partial y)}{\rho (\partial u/\partial y)^{2}}%
\thickapprox -\frac{2hg\Delta \rho }{\rho (\Delta
U)^{2}}=-\frac{4ghAt}{(\Delta U)^{2}}\text{,}
\end{equation}
where $2h$ signifies the width of the mixing layer, $\Delta \rho$ and $\Delta U$ are the differences in density and velocity, $\rho$ is the average density, $At$ is the Atwood number, $At=(\rho_{u}-\rho_{b})/(\rho_{u}+\rho_{b})=\Delta \rho/(2 \rho)$. For the cases of \ref{fig10}(c), \ref{fig10}(d), and \ref{fig10}(e), the corresponding values are obtained $R_{ic}=-0.295$, $R_{id}=-0.587$, $R_{ie}=-0.9344$. The smaller the negative $Ri$ is, the stronger the effect of shear will be. The results show a satisfactory agreement with the work by Finn \cite{Finn2014}.

Then, we focus our attention on the early stage of the RTKHI system. In the RTKHI system, RTI always plays a major role in the later stage, while the main mechanism in the early stage depends on the comparison of buoyancy and shear strength (namely, the gravity acceleration $g$ and the shear velocity $u_{0}$). That is, both types of instability develop with time, however, as there are differences between powerful and weak strength, there will be the primary one as well as secondary one. For a given gravity acceleration $g$, there is a critical shear velocity $u_C$. In the corresponding coupling system with gravity acceleration $g$ and shear velocity $u_C$, the buoyancy and shear effects are balanced. If the shear velocity $u_{0}$ of the RTKHI system is greater than $u_C$, the shearing effect in the early stage of the system is stronger, and consequently the KHI dominates in the early stage (as shown in Fig. \ref{fig9}); If the shear velocity $u_{0}$ of the RTKHI system is less than $u_C$, the shearing effect is weaker, and consequently the RTI dominates in the early stage (as shown in Fig. \ref{fig7}).

Figure \ref{fig11} shows the morphological analysis of the main mechanism in the early stage. Figures (a) and (b) are comparisons of the boundary lengths of pure RTI and different pure KHI. The small insets in the middle column are the temperature Turing patterns of pure RTI (corresponding to black curves in (a) and (b)) and the corresponding pure KHI (corresponding to green curves in (a) and (b)) at the times marked by the intersection points, P$_1$ and P$_3$ in (a) and (b).
For the convenience of description, we use the pair of typical parameters ($g$, $u_{0}$) to label the corresponding pair of systems/process.
In Figs. (a) and (b), for the pure KHI system/process, the boundary length $L$ increases with the increasing shear velocity $u_{0}$ during the initial period.
In Fig.(a), the boundary lengths of the pure RTI system with $g=0.003$ (black line) and the pure KHI system with the shear velocity $u_{0}=0.08$ (green line) keep close before the time $t=180$ at which the two curves intersect at point P$_1$, which means that, $L^{RTI}=L^{KHI}$, $Dist=\left\vert L^{RTI}-L^{KHI}\right\vert =0$, the pattern similarity $Sim=\infty$ at this moment. During the period $0 \le t \le 180$, the process similarity $Sim_P$ keeps nearly $\infty$. That is to say, the two systems/processes have the same degree of material mixing at this moment and keep nearly the same degree of material mixing before this moment from this perspective.
The $L$ curve for the pure KHI with $u_{0}=0.05$ (red line) is below the $L$ curve of the pure RTI system with $g=0.003$ (black line), and the two $L$ curves deviate more with time. That is to say, the process similarity $Sim_P$ of the pair of systems/processes decreases with time.
It is shown that the $L$ curve for the pure KHI with $u_{0}=0.1$ (blue line) and the $L$ curve for the pure RTI with $g=0.003$ (black line) intersects also at the point P$_2$. It should be pointed out that, even though $L^{RTI}=L^{KHI}$, $Dist=\left\vert L^{RTI}-L^{KHI}\right\vert =0$, the pattern similarity $Sim=\infty$ at the moment corresponding to P$_2$, but it is clear that the process similarity of this pair of systems/processes before the point p$_2$, $Sim_P$, is much less than that of the pair, ($g=0.003$, $u_{0}=0.08$ ).
Our numerical results show that, compared with the pure RTI with $g=0.003$, for the pure KHI cases, the one with $u_{0}=0.08$ is the critical case which shows the highest process similarity $Sim_P$, the larger the deviation $|u_{0}-0.08 |$, the higher the process difference $1/Sim_P$, the less the process similarity $Sim_P$. Therefore, we refer $u_{0}=0.08$ as to the critical shear velocity $u_C$ for $g=0.003$, i.e., $u_C (g=0.003) = 0.08$. This is what the point P$_5$ in Fig. (c) means.
Figure (b) shows a second set of numerical results which can be interpreted in the same way. Figure (b) shows that, compared with the pure RTI with $g=0.005$, for the pure KHI cases, the one with $u_{0}=0.1$ is the critical case which shows the highest process similarity $Sim_P$, i.e., $u_C (g=0.005) = 0.1$.  This is what the point P$_6$ in Fig. (c) means.
The other points in Fig.(c) are obtained in a similar way. It is interesting to find that $u_C$ shows a linear relationship with $g$. In the space opened by parameters $g$ and $u_C$, in the region above the line KHI dominates the RTKHI system, in the region below the line RTI dominates. If we introduce a Reynolds number $Re$ defined as
\[
Re=\rho u_C \lambda/\mu
\]
the five points in Fig.(c) will be transformed to the five points in Fig. (d), where $\lambda$ is taken as the initial wavelength of interface perturbation, the coefficient of viscosity is $\mu =\rho T/s_{v}$, and $T$ is taken as the average temperature, $T=1.0$. Figure (d) shows the parameter regions where KHI dominates and RTI dominates in the space opened by Reynolds number $Re$ and gravity $g$.

Therefore, the value of morphological total boundary length $L$ is an helpful and efficient indication for the degree of instability development or material mixing. Furthermore, $L$ of the condensed temperature field can be used to measure the ratio of buoyancy to shear strength, and consequently it can be used to quantitatively judge the main mechanism in the early stage of the RTKHI system. Specifically, when KHI (RTI) dominates, $L^{KHI} > L^{RTI}$ ($L^{KHI} < L^{RTI}$); when KHI and RTI are balanced, $L^{KHI} = L^{RTI}$.

In addition, it should be noted that the points (the intersections of the $L$ curves of pure KHI and RTI) in Figs. \ref{fig11}(a) and (b) are not the transition points of RTKHI systems. The concept of transition point only exists in the RTKHI system ($u_{0}>u_C$) in which KHI plays a major role in the early stage and RTI plays a major role in the later stage. For this type of system, although KHI plays a major role in the early stage, buoyancy always exists. The combined effect of buoyancy and shear will inevitably lead to more rapid development of instability than the pure KHI or RTI system. As a result, the transition point is earlier than the intersections of the $L$ curves in Fig. \ref{fig11}. When the gravitational acceleration is constant, the greater the shear rate, the earlier the transition point appears, as shown in Figs. \ref{fig10} (c), (d) and (e).

\subsection{Non-equilibrium characteristics of RTKHI systems}

\begin{figure}[tbp]
\center\includegraphics*%
[bbllx=10pt,bblly=90pt,bburx=568pt,bbury=710pt,width=0.5\textwidth]{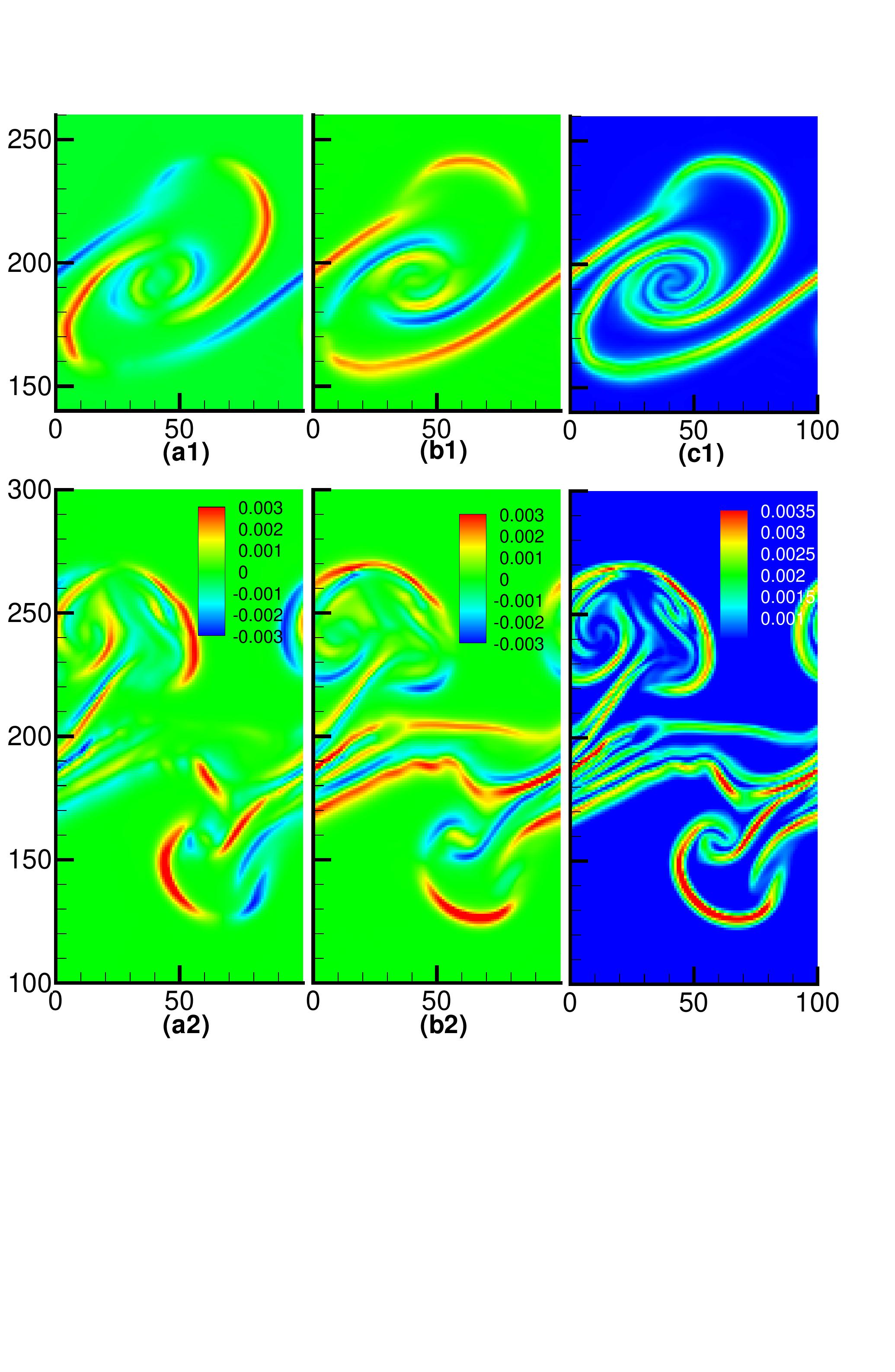}
\caption{Non-equilibrium characteristics of RTKHI system ($g=0.005$, $u_{0}=0.1$), (a) $\Delta_{(3,1)x}^{\ast}$, (b) $\Delta_{(3,1)y}^{\ast}$, and (c) $d_{(3,1)}$. The first and second lines correspond to $t=150$ and $t=250$, respectively. Each column follows the same legend.}\label{fig12}
\end{figure}

The system with hydrodynamic instability is a typical non-equilibrium flow system. Therefore, it makes sense to fully understand the various non-equilibrium behavior characteristics within such a system. In this section, the non-equilibrium characteristics of the RTKHI will be discussed. In the evolution of hydrodynamic instability, the non-equilibrium effects are significant near the interface, and basically $0$ in the position far from interface. Figure \ref{fig12} shows the contours of the non-equilibrium components $\Delta_{(3,1)x}^{\ast}$ (a), $\Delta_{(3,1)y}^{\ast}$ (b) and the corresponding NOEF strength $d_{(3,1)}$ (c) at $t=150$ and $t=250$. $\Delta_{(3,1)x}^{\ast}$ and $\Delta_{(3,1)y}^{\ast}$ correspond to the heat flux in $x$ direction and $y$ direction, respectively. From Figs. \ref{fig12}(a) and (b), it can be found that the negative heat flux and the positive heat flux appear alternately. The closer to the center of the spiral interface, the weaker of the non-equilibrium quantities. In Fig. \ref{fig12}(c), a clear double spiral structure can be seen from $d_{(3,1)}$. These information and structures cannot be found (or are not easy to be found) from the contours of temperature (as shown in Fig.\ref{fig8}(a)). Compared with the individual components, the NOEF strength $d_{(3,1)}$ provides a high resolution interface, and can be well used to describe the complete outline of interface in the RTKHI simulation.

\begin{figure*}
\begin{center}
\includegraphics[bbllx=17pt,bblly=18pt,bburx=327pt,bbury=258pt,width=0.8\textwidth]{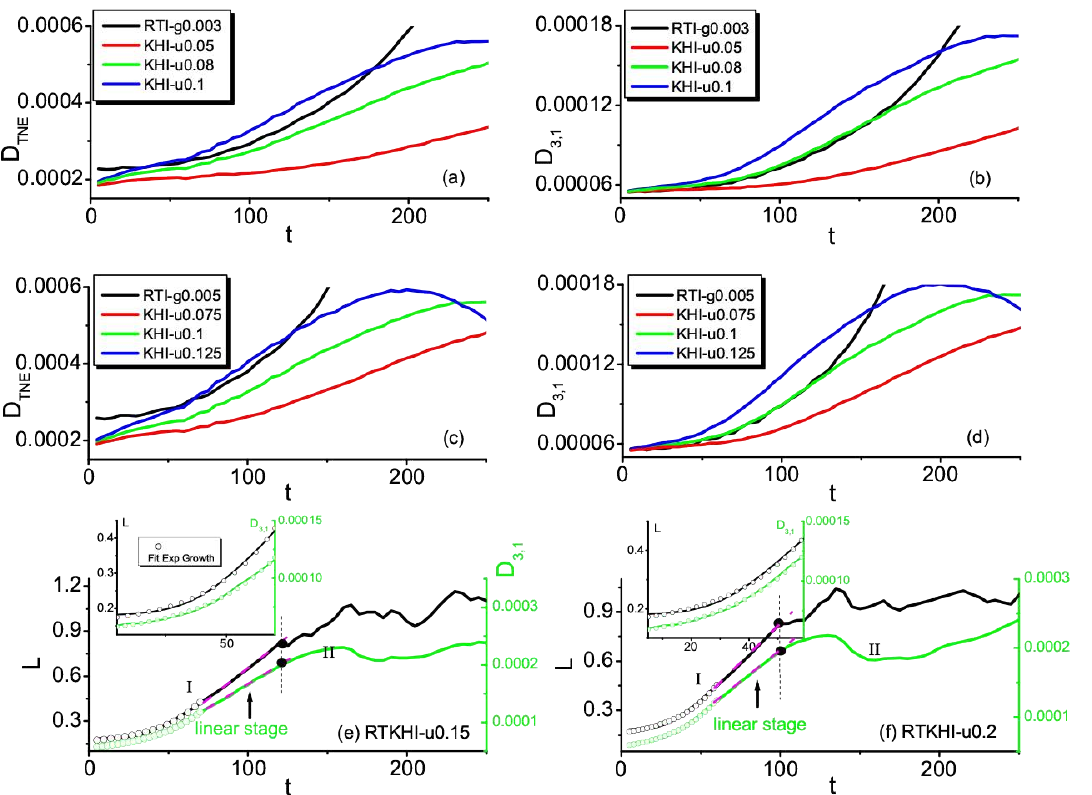}
\end{center}
\caption{The applications of non-equilibrium characteristics in the early main mechanism judgment (a)-(d) and transition point capture (e)-(f). Figs. (a) and (c) show the comparison of global average TNE strength $D_{TNE}$ between pure RTI and different pure KHI. Figs. (b) and (d) show the comparison of global average NOEF strength $D_{3,1}$ between pure RTI and different pure KHI. In Figs. (e) and (f), the evolutions of $L$ and $D_{3,1}$ are shown in the same graph for one case of RTKHI system. The shear rates in (e) and (f) are $u_{0}=0.15$ and $u_{0}=0.2$, respectively.}\label{fig13}
\end{figure*}
\begin{figure*}
\begin{center}
\includegraphics[bbllx=15pt,bblly=15pt,bburx=340pt,bbury=245pt,width=0.7\textwidth]{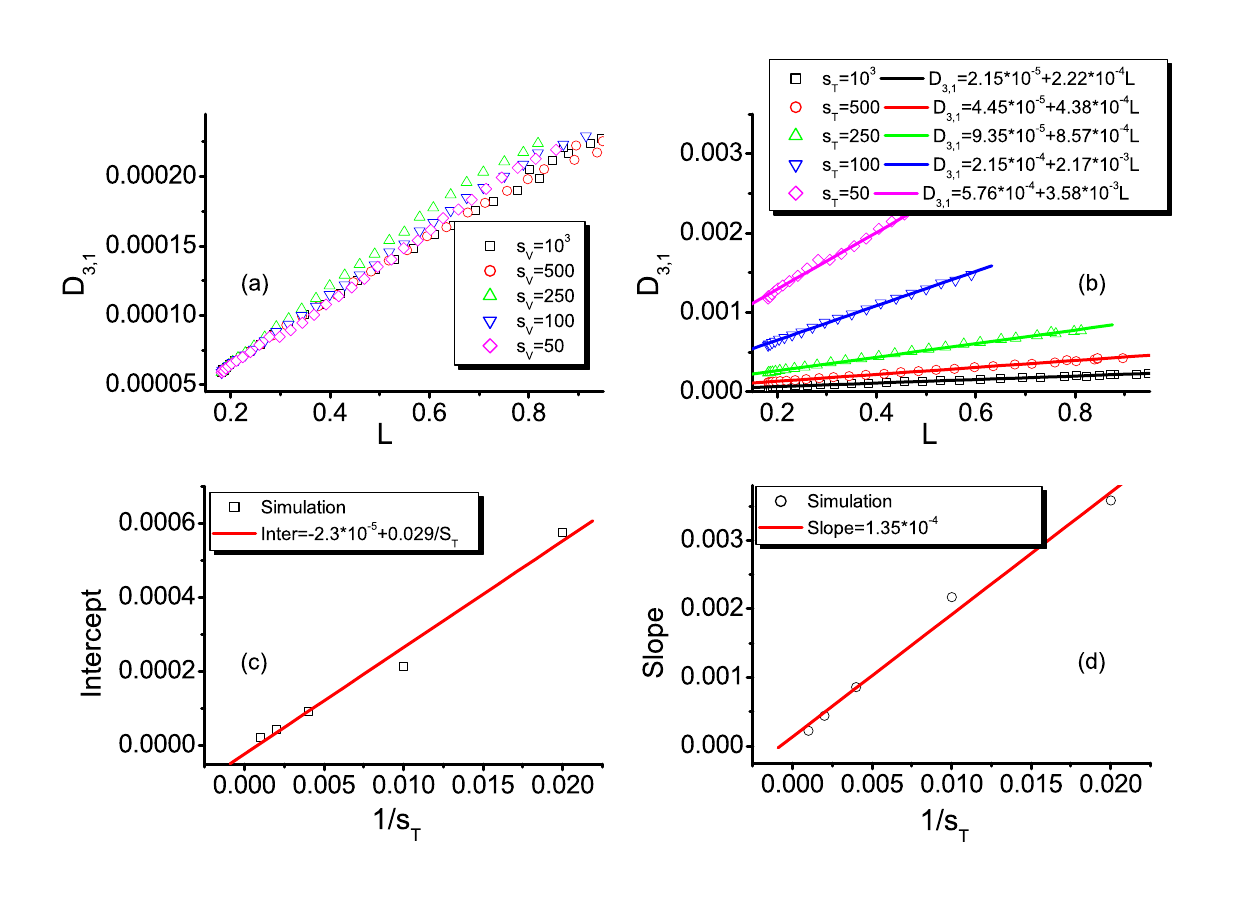}
\end{center}
\caption{The linear relationship between $L$ and $D_{3,1}$ in the early stage. Figs. (a) and (b) show the effects of viscosity and heat conductivity. The relations between intercept and slope of linear fitting shown in Fig. (b) and heat conductivity are shown in Figs. (c) and (d), respectively.}\label{fig14}
\end{figure*}

Figure \ref{fig13}(a)-(d) shows the applications of non-equilibrium characteristics in the early main mechanism judgment. Compared with the global average TNE strength $D_{TNE}$ shown in \ref{fig13}(a) and (c), the global average NOEF strength $D_{3,1}$ shown in \ref{fig13}(b) and (d) can more accurately judge the main mechanism in the early stage, and the conclusions are consistent with the morphological boundary length $L$ (as shown in figure \ref{fig11}). Specifically, when KHI (RTI) dominates in the early RTKHI system, $D_{3,1}^{KHI} > D_{3,1}^{RTI}$ ($D_{3,1}^{KHI} < D_{3,1}^{RTI}$); when KHI and RTI are balanced, $D_{3,1}^{KHI} = D_{3,1}^{RTI}$. The global average TNE strength $D_{TNE}$ cannot be used as the distinguish criterion, because the global average TNE strength is closely related to density nonuniformity, and the initial density of RTI is a function of acceleration $g$, which is completely different from the density setting of KHI.

Since morphologic quantities and non-equilibrium quantities describe the same physical process from different perspectives, there must be a relationship between them. It is worth studying which quantities are strongly correlated and which are weakly correlated. This is the core task of several of our previous work \cite{Chen2016,Chen2018}. In our previous work \cite{Chen2016,Chen2018} we have make it clear that the mean Non-Organized Energy Flux (NOEF) $D_{3,1}$ and mean temperature nonuniformity has a high correlation which is almost 1. How does the specific correlation behavior between the total boundary length $L$ and $D_{3,1}$ naturally becomes an interesting question. In figures \ref{fig13}(e) and \ref{fig13}(f), $D_{3,1}$ shows similar behavior to the boundary length $L$. We adopt the correlation function $C$ to express this similarity.
\begin{equation}
C=\frac{\overline{(L-\overline{L})(D_{3,1}-\overline{D_{3,1}})}}{\sqrt{\overline{(L-\overline{L})^{2}}\cdot
\overline{(D_{3,1}-\overline{D_{3,1}})^{2}}}}
\end{equation}
where $\overline{L}$ and $\overline{D_{3,1}}$ are the averages of $L$ and $D_{3,1}$ in the time concerned, respectively.
It is interesting to find that the two quantities, $L$ and $D_{3,1}$, always show a high correlation, especially in the early stage (This finding also exists in pure RT and pure KH systems). In figure \ref{fig13}(e) (\ref{fig13}(f)), the overall correlation degrees between $D_{3,1}$ and $L$ is approximate to $0.985$ ($0.967$), the correlation in the early stage (stage I) is $0.999$ ($0.999$),
and it decrease to $0.716$ ($0.269$) in the later stage (stage II). The physical reason is as below. $D_{3,1}$ is the mean heat flux strength, which is closely related to temperature gradient. The macroscopic quantity gradient, especially the temperature gradient, mainly exists near the interface. The greater the interface length $L$ is, the larger $D_{3,1}$ is. Therefore, $D_{3,1}$ can also be used to capture the transition points from KHI-like to RTI-like (as shown by the black vertical lines in figures \ref{fig13}(e) and \ref{fig13}(f)). When $D_{3,1}$ deviates from its constant velocity growth, the system enters the second stage. That is, the ending point of linear increasing $D_{3,1}$ can work as a physical criterion for discriminating the two stages.

The high correlation between $L$ and $D_{3,1}$ in the early stage means that $L$ and $D_{3,1}$ follows roughly a linear relationship. Figure \ref{fig14} shows the linear relationship between $L$ and $D_{3,1}$ in the early stage, where Figs. \ref{fig14}(a) and (b) are the effects of viscosity and heat conduction, respectively. The solid lines are linear fittings corresponding to the simulation results. It can be seen that heat conduction has a significant influence on the linear relationship. With the increase of heat conduction, the intercepts and slopes of the linear relationship increase approximately linearly (Figs. \ref{fig14}(c) and (d)).

\section{Conclusions}

In this paper, we investigate the coupled Rayleigh-Taylor-Kelvin-Helmholtz instability system with a multiple-relaxation time discrete Boltzmann model. To quantitatively analyse the coupled RTKHI process, we resort to morphological and non-equilibrium analysis techniques, and three cases are considered: pure RTI, pure KHI, and coupled RTKHI systems. After the initial exponential growth stage, the total boundary length $L$ of the condensed temperature field of pure KHI has an approximately constant velocity growth stage, which is different from the pure RTI. RTI always dominates in the later stage of coupled RTKHI system, while the main mechanism in the early
stage depends on the comparison of buoyancy and shear strength. Both the total boundary length $L$ of the condensed temperature field and the mean heat flux strength $D_{3,1}$ can be used to measure the ratio of buoyancy to shear strength, and to quantitatively judge the main mechanism in the early stage of the RTKHI system. Specifically, when KHI (RTI) dominates, $L^{KHI} > L^{RTI}$ ($L^{KHI} < L^{RTI}$), $D_{3,1}^{KHI} > D_{3,1}^{RTI}$ ($D_{3,1}^{KHI} < D_{3,1}^{RTI}$); when KHI and RTI are balanced, $L^{KHI} = L^{RTI}$, $D_{3,1}^{KHI} = D_{3,1}^{RTI}$, where the superscript, ``KHI (RTI) " , indicates the type of hydrodynamic instability. It is interesting to find that, (i) for the critical cases where KHI and RTI are balanced, both the critical shear velocity $u_C$ and Reynolds number $Re$ show a linear relationship with the gravity/accelaration $g$; (ii) the two quantities, $L$ and $D_{3,1}$, always show a high correlation, especially in the early stage where it is roughly $0.999$, which means that $L$ and $D_{3,1}$ follows approximately a linear relationship. The heat conduction has a significant influence on the linear relationship. For the case where the KHI dominates at earlier time and the RTI dominates at later time, the boundary length $L$ can well capture the transition point from KHI-like to RTI-like. Before the transition point of the two stages, $L^{RTKHI}$ initially increase exponentially, and then increases linearly. This linear increasing behavior ends at the transition point. Hence, the ending point of linear increasing $L^{RTKHI}$ can work as a geometric criterion for discriminating the two stages. The TNE quantity, heat flux strength $D_{3,1}^{RTKHI}$ shows similar behavior to boundary length $L$, and a strong positive correlation can be found in the early stage. Therefore, the ending point of linear increasing $D_{3,1}^{RTKHI}$ can work as a physical criterion for discriminating the two stages. The morphological boundary length $L$ is the length of the interface between light and heavy (high and low temperature) fluids. It reflects the degree of instability development and material mixing. The TNE quantity $D_{3,1}$ reflects the degree of deviation from equilibrium in different regions of the system, and can more clearly and accurately locate the position of the interface. The two criteria, $L$ and $D_{3,1}$, have different perspectives, but are consistent, have their own advantages, and complement each other. The resort to these two criteria facilitates the identification of the main mechanisms and critical time of the coupled RTKHI systems.

\section*{Acknowledgments}

FC and QZ acknowledges support from the Shandong Province Higher Educational Youth Innovation Science and Technology Program (under Grant No. 2019KJJ009).
AX acknowledges support from the National Natural Science Foundation of China (under Grant Nos. 11772064), CAEP Foundation (under Grant No. CX2019033),  the opening project
of State Key Laboratory of Explosion Science and Technology (Beijing Institute of Technology) (under Grant No. KFJJ19-01M).
YZ acknowledges support from China Postdoctoral Science Foundation (under Grant No. 2019M662521), and Scientifific Research Foundation of Zhengzhou university (under Grant No. 32211545 ).

\section*{Data Availability}

The data that support the findings of this study are available from the corresponding author upon reasonable request.

\section*{Appendix: Transformation matrix and equilibria of the kinetic moments}

The transformation matrix and the corresponding equilibrium distribution functions in KMS are constructed according to the seven moment relations. Specifically,
the transformation matrix is
\begin{equation*}
\mathbf{M}=(m_{1},m_{2},\cdots ,m_{16})^{T}\text{,}\;
\end{equation*}
\begin{equation*}
m_{1}=1\text{,}\; m_{2}=v_{ix}\text{,}\; m_{3}=v_{iy}\text{,}\;
m_{4}=(v_{i\alpha }^{2}+\eta_{i}^{2})/2\text{,}\;
\end{equation*}
\begin{equation*}
m_{5}=v_{ix}^{2}\text{,}\; m_{6}=v_{ix}v_{iy}\text{,}\;
m_{7}=v_{iy}^{2}\text{,}\;
\end{equation*}
\begin{equation*}
m_{8}=(v_{i\beta}^{2}+\eta
_{i}^{2})v_{ix}/2\text{,}\; m_{9}=(v_{i\beta
}^{2}+\eta_{i}^{2})v_{iy}/2\text{,}\;
\end{equation*}
\begin{equation*}
m_{10}=v_{ix}^{3}\text{,}\; m_{11}=v_{ix}^{2}v_{iy}\text{,}\;
m_{12}=v_{ix}v_{iy}^{2}\text{,}\; m_{13}=v_{iy}^{3}\text{,}\;
\end{equation*}
\begin{equation*}
m_{14}=(v_{i\chi}^{2}+\eta _{i}^{2})v_{ix}^{2}/2\text{,}\;
m_{15}=(v_{i\chi }^{2}+\eta_{i}^{2})v_{ix}v_{iy}/2\text{,}\;
\end{equation*}
\begin{equation*}
m_{16}=(v_{i\chi }^{2}+\eta_{i}^{2})v_{iy}^{2}/2\text{.}\;
\end{equation*}

The corresponding equilibrium distribution functions in KMS are
\begin{equation*}
\hat{f}_{1}^{eq}=\rho \text{,}\; \hat{f}_{2}^{eq}=\rho
u_{x}\text{,}\; \hat{f}_{3}^{eq}=\rho u_{y}\text{,}\;
\hat{f}_{4}^{eq}=e\text{,}\;
\end{equation*}
\begin{equation*}
\hat{f}_{5}^{eq}=P+\rho u_{x}^{2}\text{,}\; \hat{f}_{6}^{eq}=\rho
u_{x}u_{y}\text{,}\; \hat{f}_{7}^{eq}=P+\rho u_{y}^{2}\text{,}\;
\end{equation*}
\begin{equation*}
\hat{f}_{8}^{eq}=(e+P)u_{x} \text{,}\;
\hat{f}_{9}^{eq}=(e+P)u_{y}\text{,}\; \hat{f}_{10}^{eq}=\rho
u_{x}(3T+u_{x}^{2})\text{,}\;
\end{equation*}
\begin{equation*}
\hat{f}_{11}^{eq}=\rho u_{y}(T+u_{x}^{2})\text{,}\;
\hat{f}_{12}^{eq}=\rho u_{x}(T+u_{y}^{2})\text{,}\;
\hat{f}_{13}^{eq}=\rho u_{y}(3T+u_{y}^{2})\text{,}\;
\end{equation*}
\begin{equation*}
\hat{f}_{14}^{eq}=(e+P)T+(e+2P)u_{x}^{2}\text{,}\;
\hat{f}_{15}^{eq}=(e+2P)u_{x}u_{y}\text{,}
\end{equation*}
\begin{equation*}
\hat{f}_{16}^{eq}=(e+P)T+(e+2P)u_{y}^{2}\text{,}
\end{equation*}
where pressure $P=\rho RT$ and energy $ e=b\rho RT/2+\rho u_{\alpha }^{2}/2$. $R$\ is the specific gas constant and $b$ is a constant related to the specific-heat-ratio $\gamma$ by $\gamma=(b+2)/b$.

Replacing $v_{i\alpha }$ by $v_{i\alpha }-u_{\alpha }$ in the transformation matrix $\mathbf{M}$, matrix $\mathbf{M}^{\ast }$ is expressed as follows:
\begin{equation*}
\mathbf{M}^{\ast }=(m_{1}^{\ast },m_{2}^{\ast },\cdots ,m_{16}^{\ast })^{T},
\end{equation*}
\begin{equation*}
m_{1}^{\ast }=1\text{,}\; m_{2}^{\ast }=v_{ix}-u_{x}\text{,}\; m_{3}^{\ast }=v_{iy}-u_{y}\text{,}
\end{equation*}%
\begin{equation*}
m_{4}^{\ast }=((v_{ix}-u_{x})^{2}+(v_{iy}-u_{y})^{2}+\eta _{i}^{2})/2\text{,}
\end{equation*}%
\begin{equation*}
m_{5}^{\ast }=(v_{ix}-u_{x})^{2}\text{,}\; m_{6}^{\ast }=(v_{ix}-u_{x})(v_{iy}-u_{y})\text{,}\;
m_{7}^{\ast }=(v_{iy}-u_{y})^{2}\text{,}
\end{equation*}%
\begin{equation*}
m_{8}^{\ast }=[(v_{ix}-u_{x})^{2}+(v_{iy}-u_{y})^{2}+\eta
_{i}^{2}](v_{ix}-u_{x})/2\text{,}
\end{equation*}%
\begin{equation*}
m_{9}^{\ast }=[(v_{ix}-u_{x})^{2}+(v_{iy}-u_{y})^{2}+\eta
_{i}^{2}](v_{iy}-u_{y})/2\text{,}
\end{equation*}%
\begin{equation*}
m_{10}^{\ast }=(v_{ix}-u_{x})^{3}\text{,}\;
m_{11}^{\ast }=(v_{ix}-u_{x})^{2}(v_{iy}-u_{y})\text{,}
\end{equation*}%
\begin{equation*}
m_{12}^{\ast }=(v_{ix}-u_{x})(v_{iy}-u_{y})^{2}\text{,}\;
m_{13}^{\ast }=(v_{iy}-u_{y})^{3}\text{,}
\end{equation*}%
\begin{equation*}
m_{14}^{\ast }=[(v_{ix}-u_{x})^{2}+(v_{iy}-u_{y})^{2}+\eta
_{i}^{2}](v_{ix}-u_{x})^{2}/2\text{,}
\end{equation*}%
\begin{equation*}
m_{15}^{\ast }=[(v_{ix}-u_{x})^{2}+(v_{iy}-u_{y})^{2}+\eta
_{i}^{2}](v_{ix}-u_{x})(v_{iy}-u_{y})/2\text{,}
\end{equation*}%
\begin{equation*}
m_{16}^{\ast }=[(v_{ix}-u_{x})^{2}+(v_{iy}-u_{y})^{2}+\eta
_{i}^{2}](v_{iy}-u_{y})^{2}/2.
\end{equation*}


\section*{References}

\bibliography{POF20-AR-02364-REF}

\end{document}